# Water-Superstructured Solid Fuel Cells

Wei Zhang[1], Siyuan Fang[1], Hanrui Su[1], Wei Hao[2], Bohak Yoon[2], Gyeong S. Hwang[2], Kai Sun[3], Chung-Fu Chen[4], Yu Zhu[4], Yun Hang Hu[1]*

**Affiliations:**

[1]Department of Materials Science and Engineering, Michigan Technological University; Houghton, Michigan, 49931, United States.

[2]McKetta Department of Chemical Engineering, University of Texas at Austin; Austin, Texas, 78712, United States.

[3]Department of Materials Science and Engineering, University of Michigan; Ann Arbor, Michigan, 48109, United States.

[4]School of Polymer Science and Polymer Engineering, The University of Akron; Akron, Ohio, 44325, United States.

*Corresponding author. Email: yunhangh@mtu.edu

**Abstract:** Protonic ceramic fuel cells can be operated at low temperatures, but their performances relying on bulk ion transfer in solid electrolytes are usually limited by much lower proton conductivity than 0.1 S $cm^{-1}$ below 600 ºC. Herein, however, we report a strategy for $Al_2O_3$ insulator to become a protonic superconductor, namely, in-situ generation of superstructured-water in porous $Al_2O_3$ layer could realize the unprecedented water-mediated proton transfer on $Al_2O_3$ surface, attaining ultrahigh proton conductivity of 0.13 S $cm^{-1}$ at 550 ºC. With such a water-superstructured proton-superconductor, we created water-superstructured solid fuel cell, achieving very high power density of 1036 mW $cm^{-2}$ and high open circuit voltage above 1.1 V at 550 ºC with $H_2$ fuel. This provides a general approach to develop protonic superconductors and solid fuel cells.

## Introduction

Fuel cells, which are an important type of energy devices, can overcome the limitations of fuel combustion efficiency. Compared with proton exchange membrane fuel cells that require expensive noble metal catalysts with poor tolerance to gas impurity, ceramic fuel cells offer a superior capability for power generation with fuel flexibility, high efficiency, and non-noble metal catalysts (*1–11*). The yttria-stabilized zirconia (YSZ)-based solid oxide fuel cells (SOFCs) represent the 1st generation ceramic cells, but their high operating temperature up to 1000 °C caused material compatibility issues and high cost (*12*). The development of new oxide-ion conducting electrolytes, such as gadolinium-doped ceria (GDC) and samarium-doped ceria (SDC), created the 2nd generation SOFCs with a decreased operating temperature of 600-700 ºC (*13, 14*). The 3rd generation nanostructured SOFCs can be operated even at 450-600 ºC, but their performance declines rapidly with decreasing temperature owing to the high activation energy of oxide-ion conduction (*4, 14*). In contrast, protonic conduction in proton-conductive oxides, such as yttrium-doped barium zirconate (BZY), possesses a decreased activation energy, enabling protonic ceramic fuel cells (PCFCs) to be operated at a lower temperature and thus improving the sealing, durability, and cost (*15–19*). Nevertheless, conductivities of typical proton ceramic electrolytes are still much smaller than 0.1 S $cm^{-1}$ at a relatively low temperature (< 600 ºC) (*18, 20, 21*), leading to much worse performance of PCFCs than that of high-temperature SOFCs and thus restricting the commercial application of PCFCs. Two technical approaches are being exploited to improve the proton conductivity for PCFCs. One is to reduce the electrolyte thickness, thus decreasing the proton transfer distance (*7, 21*). However, fabricating an ultrathin electrolyte film requires advanced techniques that unavoidably make mass production difficult and costly. Furthermore, it may cause "short circuits" due to electron conduction if the electrolyte film is too thin to ensure the isolation between the anode and the cathode. Another route is to modify current ceramic proton conductors and invent new materials, which is strongly determined by the progress of materials science and engineering (*7, 21, 22*).

As a breakthrough, Mavrikakis, Besenbacher and co-workers directly observed water-mediated proton diffusion on FeO surface by high-speed scanning tunneling microscopy (STM) (*23*). Furthermore, protons and water were found to be retained in the grain boundary or "internal surface" of nanograined YSZ, contributing to proton conductivity (*24*). The necessity of adsorbed water for fast proton conduction on sulfated zirconia was also reported (*25*). These suggest that the surface transfer of proton on a metal oxide is dependent on water-promotion instead of the intrinsic ionic bulk conductivity of the metal oxide. Therefore, as a hypothesis, we propose that the metal oxide ionic-insulator with condensed/adsorbed water (namely, superstructured water on a solid surface) would become a proton conductor via water-mediate proton surface diffusion. Because $\gamma$-$Al_2O_3$ is an ionic-insulator with excellent chemical and thermal stability (*26*) and remarkable affinity for water (*27*), we selected it to test this hypothesis. It was demonstrated that the $Al_2O_3$ can be changed from a ceramic insulator to a protonic superconductor by in-situ forming the superstructured-water in a porous $Al_2O_3$ layer for surface water-mediated proton transfer. Furthermore, with such a water-superstructured porous $Al_2O_3$ layer for fast proton transfer between the anode and the cathode, we created a new type of solid fuel cell—water-superstructured solid fuel cell (WSSFC), achieving a very high power output of 1036 mW $cm^{-2}$ and high open circuit voltage (OCV) above 1.1 V when operated with $H_2$ as the fuel at 550 ºC.

## Results and discussion

### *Design of WSSFC based on modeling and experimental observations*

Water-mediated proton diffusion on metal oxide surfaces was experimental demonstrated (*23-25*), which has stimulated us to propose that water would be able to promote the surface transfer of protons on a metal oxide ionic-insulator. The feasibility of this hypothesis was examined by our density functional theory (DFT) calculations with $Al_2O_3$ as follows. First, a $H_2O$ molecule was added to the (100) surface of $\gamma$-$Al_2O_3$ to form an adsorbed $H_2O$. Meanwhile, a proton was introduced to the $Al_2O_3$ and optimized to the stable state, at which the proton formed hydroxy groups with the O atoms bonded to the fourfold coordinated Al. The proton transfer on the surface of $\gamma$-$Al_2O_3$ with assistance of the adsorbed $H_2O$ molecule was simulated based on the climbing image nudged elastic band (ci-NEB) method. The proton from the hydroxy group on the surface of $\gamma$-$Al_2O_3$ contacts the adsorbed $H_2O$ molecule to generate the $H_3O^+$ transition state, which elongates and then breaks the O-H bond of the $H_2O$ molecule to give a proton to the surface of $\gamma$-$Al_2O_3$ and thus form another hydroxy group (**Fig. 1A**). Overall, the process could be considered as a proton transfer from one to another O atom bonded with the fourfold coordinated Al with the assistance of the $H_2O$ molecule via the $H_3O^+$ transition state. The energy barrier of the whole proton transfer process is 0.394 eV, which is much lower than that (1.019-1.215 eV) without water-assistance (**fig. S1**). Furthermore, to reveal the effect of intermolecular hydrogen bonding between $H_2O$ molecules on the proton transfer on the $Al_2O_3$ surface, the adsorbed $H_2O$ molecule was replaced by a water monolayer (ML) consisting of 10 $H_2O$ molecules (with intermolecular hydrogen bonding) on the surface of γ-$Al_2O_3$. **Fig. S2** shows the similar water-mediated proton transfer approach on the surface of $Al_2O_3$ via the $H_3O^+$ transition state, and its proton transfer energy barrier (0.303 eV) is even lower than that (0.394 eV) with an adsorbed $H_2O$. Therefore, the energy barrier of proton transfer promoted either with the water ML or the adsorbed water molecule is lower than the activation barriers (0.4-0.6 eV) of reported excellent proton ceramic conductors (*28*). Furthermore, the energy barrier of water-mediated proton transfer is also smaller than the adsorption energy (0.514 eV) of ML $H_2O$ molecule and that (0.642-0.813 eV) of independent $H_2O$ molecule on $Al_2O_3$. These clearly demonstrate the feasibility of the fast water-mediated proton transfer on the surface of γ-$Al_2O_3$.

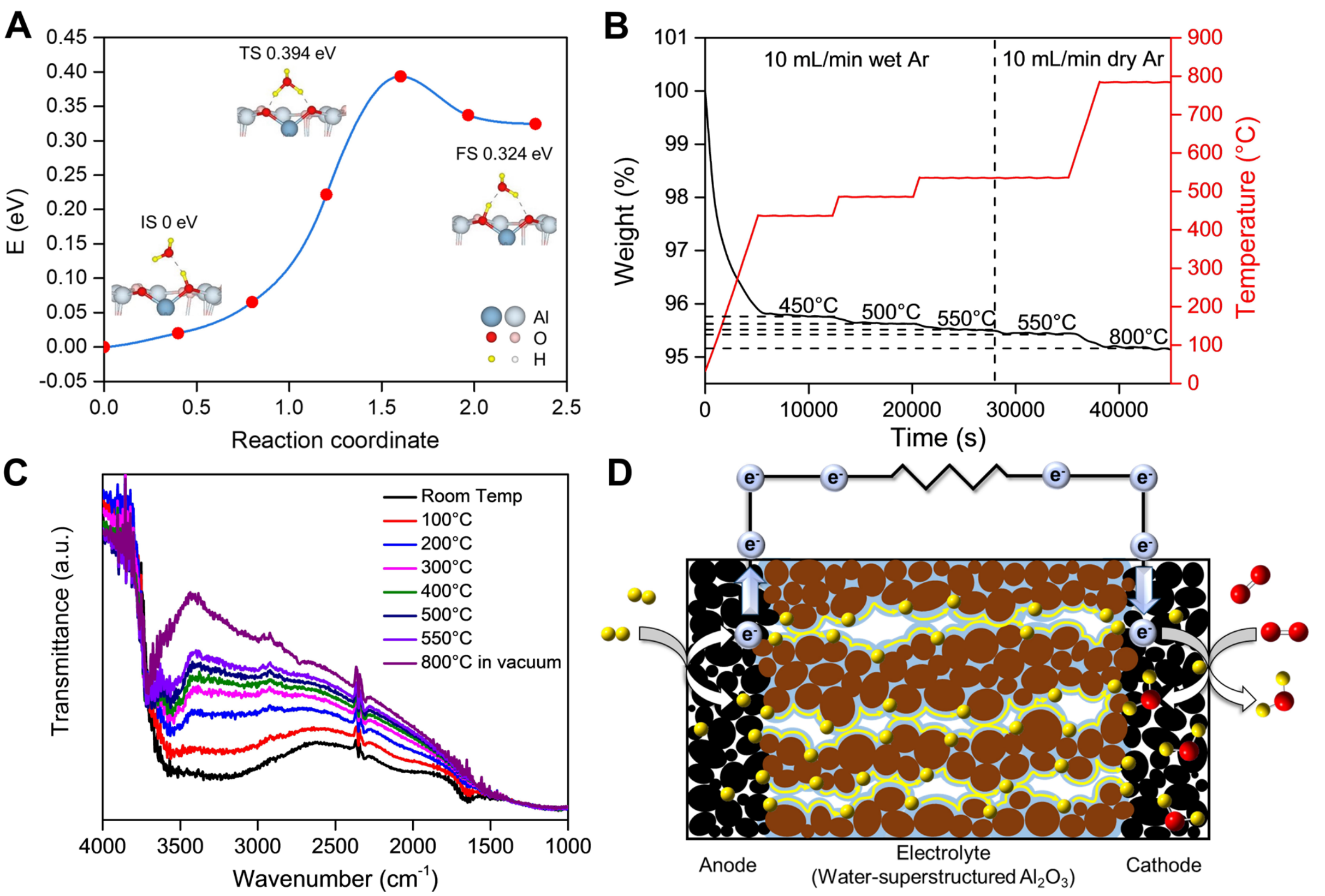

**Fig. 1. Design of water-superstructured solid fuel cell (WSSFC)**. **(A)** Water-mediated proton transfer on $Al_2O_3$ surface with an adsorbed $H_2O$ molecule demonstrated by DFT calculation. **(B)** Thermogravimetric curve of $Al_2O_3$ pellet (pressed at 100 MPa) in 3% $H_2O$/Ar atmosphere. **(C)** In-situ FT-IR spectra of $Al_2O_3$ at various temperatures in 3% $H_2O$/Ar atmosphere. **(D)** Schematic design of WSSFC (black for electrode material, brown for $Al_2O_3$, light blue for water, red for oxygen atom, and yellow for proton or hydrogen atom).

Water on surface of $Al_2O_3$ (with external surface area of 19 $m^2\ g^{-1}$ and total BET surface area of 113 $m^2\ g^{-1}$, **figs. S3-S5**) under $H_2O$/Ar atmosphere was evaluated by thermogravimetric (TG) analysis at elevated temperatures (**Fig. 1A** and **tables S1 and S2**). Although water on the $Al_2O_3$ pellet (pressed at 100 MPa) decreased with increasing temperature (**Fig. 1B**), 4.05 mg $g^{-1}$of water still retained on the $Al_2O_3$ surface in the 3% $H_2O$/Ar atmosphere even at 550 °C (**table S1**). When $H_2O$ content in Ar atmosphere increased to 9%, water on the $Al_2O_3$ at 550 °C increased to 9.01 mg $g^{-1}$, which could cover 6 $m^2$ $g^{-1}$ surface area of $Al_2O_3$ that is 32% of its external surface (19 $m^2$ $g^{-1}$) (**table S2**). Furthermore, the water retained on $Al_2O_3$ was further supported by the in-situ diffuse reflectance infrared Fourier transform spectrum, namely, the strong and broad IR band between 1800-3800 $cm^{-1}$ was observed in the 3% $H_2O$/Ar atmosphere from room temperature to elevated temperatures (**Fig. 1C**). The broad IR band, which was widely used to identify the water on the surface of metal oxides (*24, 29*), further confirmed the existence of water on $Al_2O_3$ even at 550 °C. Such water on $Al_2O_3$ surface at an elevated temperature is the so-called superstructured-water.

From the above modeling and experimental observations, we propose a new type of fuel cell—water-superstructured solid fuel cell (WSSFC), in which a porous $Al_2O_3$ film with adsorbed water molecules as a water-superstructured electrolyte for fast surface water-mediated proton transfer from the anode to the cathode (**Fig. 1D**). The superstructured-water in porous $Al_2O_3$ film is in-situ formed and remained by the water continuously produced during the cell operation.

### *Fabrication, evaluation, and characterization of WSSFCs*

Based on the above design, we fabricated the WSSFC with a simple one-step method: The powder with a sandwich distribution ($LiNi_{0.8}Co_{0.15}Al_{0.05}O_2$, $\gamma$-$Al_2O_3$, and $LiNi_{0.8}Co_{0.15}Al_{0.05}O_2$) was physically pressed with a pressure of 100 MPa at room temperature into three porous layers as a symmetric cell, in which the $\gamma$-$Al_2O_3$ layer was located between two $LiNi_{0.8}Co_{0.15}Al_{0.05}O_2$ (LNCA) electrodes. The sandwich structure of the cell was confirmed by scanning electron microscope (SEM) images (**Fig. 2A**), namely, the $Al_2O_3$ layer with about 0.3 mm thickness is between two LNCA electrode layers. The porous structure of the cell disk was quantitatively analyzed by X-ray computed tomography (CT) technique, namely, the porosities of two LNCA electrodes are 27-30%, while the layer of $\gamma$-$Al_2O_3$ possesses the porosity of about 26%, which are consistent with SEM measurements (**table S3**). The pore size distribution of the $Al_2O_3$ layer from $N_2$ adsorption-desorption isotherm measurement demonstrates the rich mesopores mainly in 2-40 nm (**fig. S6**), even though some macropores in the $Al_2O_3$ layer were also observed in the SEM image (**Fig. 2A**). In contrast, macropores at μm level are dominant in the electrodes (**Fig. 2A**). Furthermore, the constitution of continuous channels from these pores in the $Al_2O_3$ layer was revealed by gas analysis, namely, $H_2$ fed to the anode was detected by on-line gas chromatography (GC) in the cathode side without cell operation at room temperature (**fig. S7A**). To test the electrochemical performance of the cell, $H_2$ and air flows were introduced into the anode and reduced cathode of the cell, respectively. At the beginning, $H_2$ and air react with each other due to their diffusion through the continuous channels and produce water at 550 °C. The produced water diffused into porous $Al_2O_3$ layer to form adsorbed water as the superstructured-water through the porous structure of the $Al_2O_3$ layer, creating a proton superconductive electrolyte (See more discussion in the next section). Such a water-superstructured cell exhibited excellent performance as shown by the polarization (I-V) and corresponding power output (I-P) curves at low operating temperatures of 450-550 °C (**Fig. 2B**). The high OCV values above 1.1 V were obtained. Furthermore, the device achieved the very high peak power density of 1036 mW $cm^{-2}$ at 550 °C, which is much larger than those (below 600 mW $cm^{-2}$) of most reported PCFCs (*28, 30, 31*). Even when the operating temperature decreased to 525, 500, 475, and 450 °C, the high power densities of 920, 793, 631, and 454 mW $cm^{-2}$ were still obtained, respectively. The cell also exhibited excellent stability. Although the cell showed slight performance decrease with operation time at 550 ºC, its performance remained constant without degradation for 100 hours at 500 ºC (**Fig. 2C**), which is comparable to the best stability of reported SOFCs and PCFCs (*30, 31*).

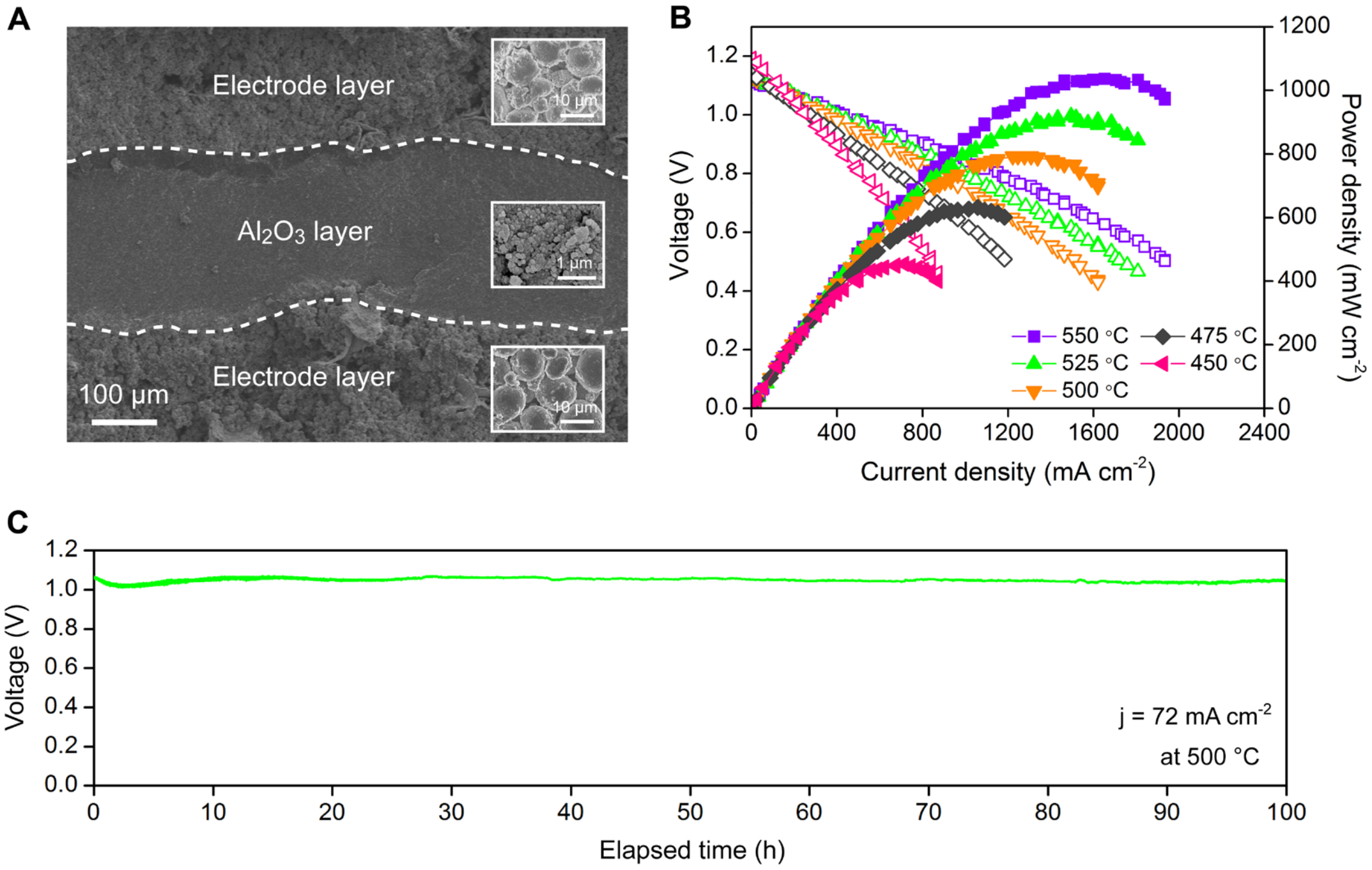

**Fig. 2. Structure and electrochemical performance of water-superstructured solid fuel cell (WSSFC). (A)** Cross-section SEM image of the fabricated cell with porous $Al_2O_3$ layer between two porous $LiNi_{0.8}Co_{0.15}Al_{0.05}O_2$ electrodes. **(B)** I-V and I-P curves of the WSSFC with $H_2$ fuel at 550–450 °C. **(C)** Stability evaluation at a constant current density at 500 °C for 100 hours.

### *Cell disk characterization and ion transfer evaluation of WSSFCs*

The high OCV value above 1.1 V reveals the negligible electronic and mechanical leakage of the cell (**Fig. 2B**). Its negligible gas leakage was further confirmed by gas analysis (**fig. S7B**). In contrast, when we on-purpose drilled a hole (1 mm diameter) through the cell disk and thus created gas leakage (**fig. S8**), OCV decreased to almost zero (0.02 V). Therefore, the high OCV indicates the in-situ formation of a plugger in the porous cell disk to block gas diffusion between electrodes during the operation. This would be related to the LNCA electrodes because the replacement of LNCA with other electrodes resulted in zero OCV (**table S4**). Indeed, SEM images showed the denser interface (about 10 μm) between the anode and the $Al_2O_3$ layer (**Fig. 3A**) and the denser interface (about 8 μm) between the cathode and the $Al_2O_3$ layer (**Fig. 3B**). Compared to the electrodes and $Al_2O_3$ layer, the interfaces exhibited about 50% decrease in porosity (**table S5**). The in-situ produced water would easily plug narrower channels of the denser interfaces to block $H_2$ and $O_2$ gas diffusion in the cell disk during its operation. As a result, the direct reaction between $H_2$ and $O_2$ was greatly inhibited, and $H_2$ could be activated to generate protons and electrons at the anode in the cell. The produced electrons transfer through an external circuit to reach the cathode and reduce $O_2$ molecules into $O^{2-}$ ions, while the generated protons fast transfer through the water-superstructured porous $Al_2O_3$ layer to react with $O^{2-}$ ions at the $Al_2O_3$-cathode interface with the formation of water. Consequently, excellent performance with the high OCV value above 1.1 V was obtained for the cell.

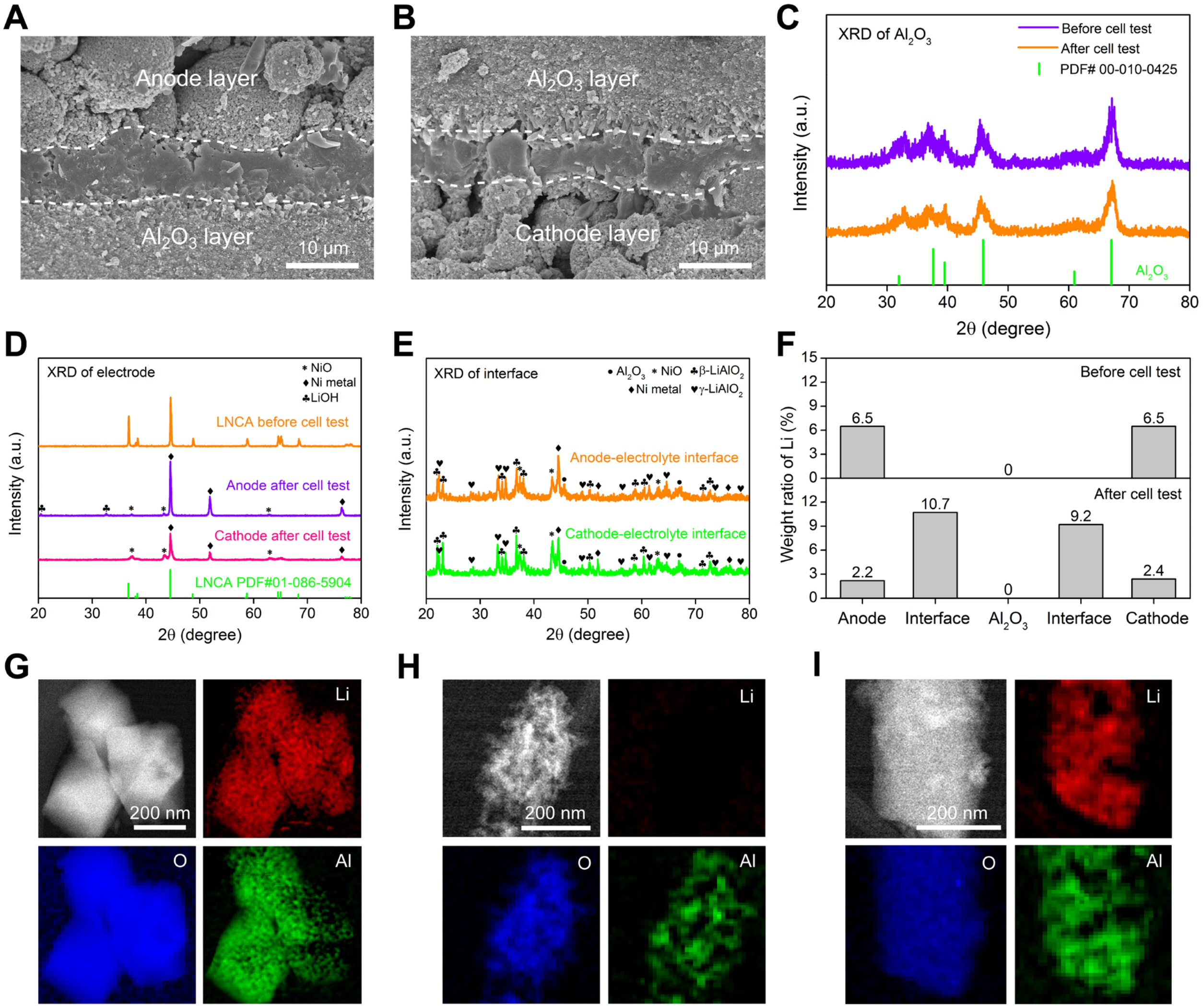

**Fig. 3. Material characterization of WSSFC.** SEM images of **(A)** anode-$Al_2O_3$ interface and **(B)** cathode-$Al_2O_3$ interface after cell test. XRD patterns of **(C)** $Al_2O_3$ layer, **(D)** electrodes, and **(E)** electrode-$Al_2O_3$ interfaces. **(F)** Weight ratio of Li in the cell materials from ICP-OES measurement. STEM-EELS elemental distributions of **(G)** anode-$Al_2O_3$ interface, **(H)** $Al_2O_3$ layer, and **(I)** cathode-$Al_2O_3$ interface after cell test.

The materials of the tested cell were deeply evaluated. X-ray diffraction (XRD) patterns revealed that the γ-$Al_2O_3$ crystal structure (PDF# 00-010-0425) of the $Al_2O_3$ layer remained unchanged (**Fig. 3C**), whereas the LNCA electrodes decomposed into Ni and NiO with strong peak intensity and LiOH with weak peak intensity after cell operation at 550 ºC (**Fig. 3D**). The decomposition of LNCA electrodes was also reported for SOFCs with ion conductive electrolytes (such as $SrTiO_3$ and $Gd_{0.1}Ce_{0.9}O_{1.95}$) after cell operation with $H_2$ fuel at an elevated temperature (*32*–*34*). Furthermore, β-$LiAlO_2$ and γ-$LiAlO_2$ were detected by XRD with low crystalline in the two interfaces due to the reaction between LiOH (from the decomposition of LNCA) and $Al_2O_3$ (**Fig. 3E**), and such a reaction could inhibit $Li^+$ diffusion into the $Al_2O_3$ layer. The inductively coupled plasma (ICP) element analysis revealed the distribution of Li in the tested cell disk, namely, the Li contents of 2.2~2.4 wt% in the anode and cathode and about 9~10 wt% in the two interfaces, but zero in the middle $Al_2O_3$ layer (**Fig. 3F**). The Li state in the anode and cathode was identified as LiOH by X-ray photoelectron spectroscopy (XPS) (**fig. S9**). The existence of Li in the interfaces and its absence in the middle $Al_2O_3$ layer were further confirmed by electron energy-loss spectroscopy (EELS) mappings (**Figs. 3G, 3H, and 3I**) and XPS spectra (**fig. S10**). Therefore, the contribution of any Li ion-based compounds (such as LiOH, generated from the decomposition of LNCA electrodes) to the ionic conductivity of the $Al_2O_3$ layer is excluded in the cell. However, LiOH plays an important role in the cathode (**table S6** and **fig. S11** with detail discussion in supplementary material).

The excellent performance of the WSSFC was further supported by the following characterization of cell properties. The impedance plot shows the very low electrolyte resistance for the water-superstructured porous $Al_2O_3$ layer (**Fig. 4A**), leading to ultrahigh proton conductivity of 0.13 S $cm^{-1}$ at 550 °C (**table S7**), which is much higher than reported values of proton conductors at the same temperature (*28, 30, 31*). The high proton conductivities of 0.11, 0.09, 0.08, and 0.07 S $cm^{-1}$ were also obtained at 525, 500, 475, and 450 °C, respectively (**Fig. 4B**, **table S7**, and **fig. S13**). Although the intrinsic $Al_2O_3$ is a well-known ionic insulator, the water-superstructured porous $Al_2O_3$ layer is responsible for the fast proton transfer. This was supported by the in-situ TG and FTIR analysis, which demonstrated the appreciable amount of water on $Al_2O_3$ surface (as super-structured water) in the $H_2O$-containing atmosphere at elevated temperatures up to 550 °C (**Figs. 1B and 1C**). Furthermore, the correlation between the proton conductivity and the operating temperature of the water-superstructured porous $Al_2O_3$ layer revealed a small activation energy of 0.343 eV for proton transfer (**Fig. 4B**), which is well consistent with the energy barriers (0.303~0.394 eV) of $H_2O$-mediated proton transfer on $Al_2O_3$ surface from the DFT calculations (**Fig. 1A** and **fig. S2**). In this proton transfer mechanism, the proton exchange between the surface OH group of $Al_2O_3$ and the $H_2O$ molecule via the $H_3O^+$ transition state plays an important role. To experimentally confirm the proton exchange, liquid $H_2O$ was dropped on $Al_2O_3$ to ensure the sufficient formation of surface OH groups at room temperature, followed by dropping liquid $D_2O$ and then increasing temperature. As shown in **Fig. 4C**, one can see the broad IR band of condensed $H_2O$ between 2800 and 3800 $cm^{-1}$ and the broad one of condensed $D_2O$ between 1700 and 2500 $cm^{-1}$. Furthermore, the stretching band (centered at 2730-2770 $cm^{-1}$) of OD group of $Al_2O_3$ started to occur at 200 °C and its IR intensity increased with increasing temperature, clearly demonstrating the D-H exchange between OH of $Al_2O_3$ and D of $D_2O$. The participation of OH groups of $Al_2O_3$ in the proton transfer was further supported by XPS spectra, namely, the content of surface OH group in the $Al_2O_3$ increased from 21.3% to 44.3% after the cell operation (**Fig. 4D**), indicating the bonding of protons to the surface O atoms of $Al_2O_3$. These provide the strong evidences for the water-mediated proton transfer via the $H_3O^+$ transition state in the water-superstructured $Al_2O_3$ layer at elevated temperatures. This transition state could be further supported by the reported observation, in which $H_3O^+$-like transition state was found in the water-mediated diffusion of proton on a FeO film by STM at room temperature or below (*23*), which revealed its fundamental difference from the Grötthus mechanism that a proton transfers via a "water wire" in the liquid water phase (*35*). The high stability of the $\gamma$-$Al_2O_3$ layer in the water-superstructured cell was demonstrated by XPS spectra, XRD patterns, and TEM images, namely, the Al valence (**Fig. 4E**), the crystal structure (**Fig. 3C**), and the particle shape and size (25 nm) (**Fig. 4F**) of the $\gamma$-$Al_2O_3$ layer remained unchanged after the cell operation.

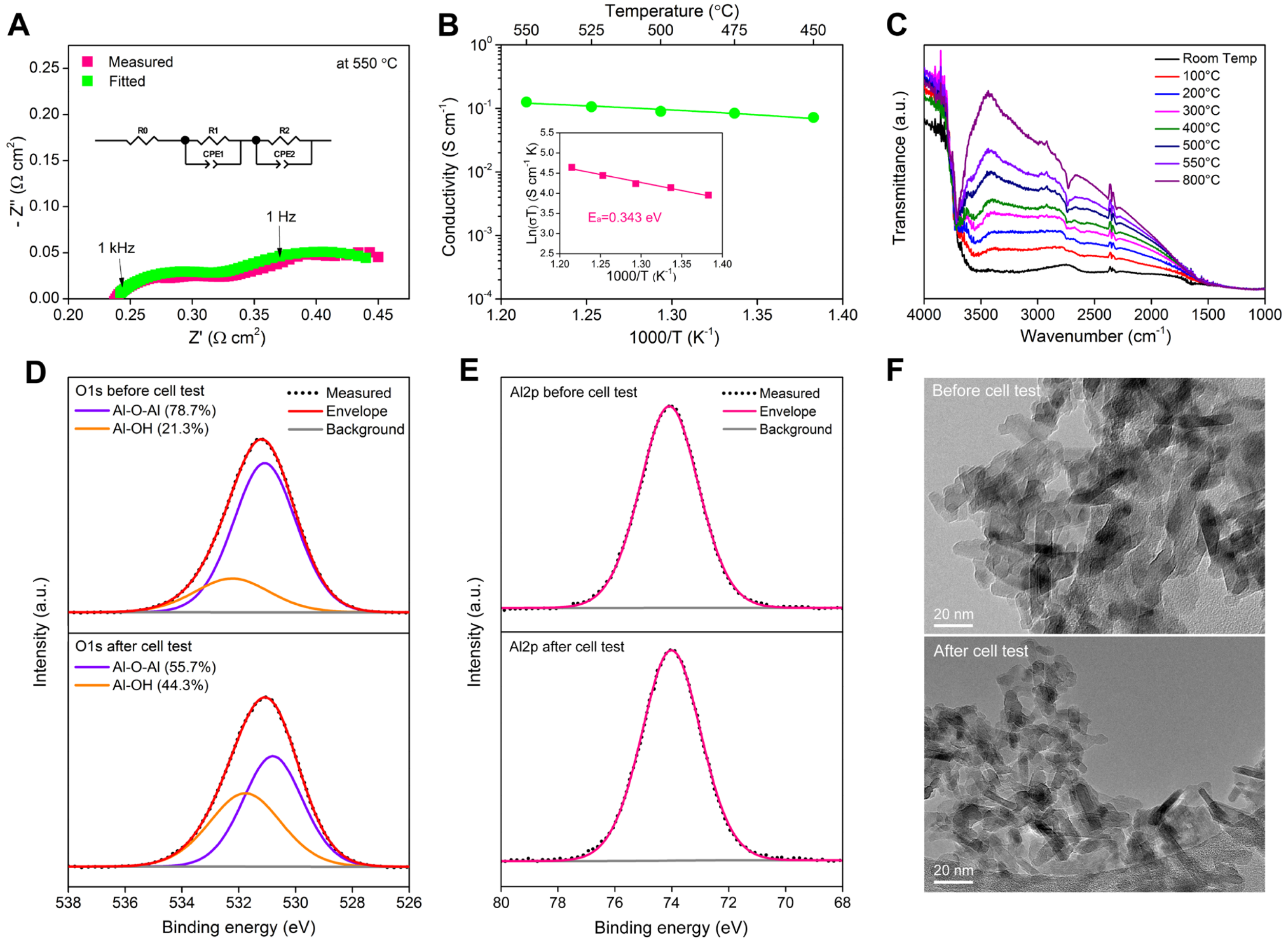

**Fig. 4. Characterization of water-superstructured $Al_2O_3$ layer**. **(A)** Nyquist plots of the water-superstructured porous $Al_2O_3$-based fuel cell under $H_2$/air open-circuit conditions. **(B)** Proton conductivity of water-superstructured porous $Al_2O_3$ layer vs. temperature. **(C)** In-situ FT-IR spectra of condensed $H_2O$ and $D_2O$ over $Al_2O_3$ at various temperatures. **(D)** $O_{1S}$ XPS spectra of $Al_2O_3$ before and after the cell test at 550 ºC. **(E)** $Al_{2P}$ XPS spectra of $Al_2O_3$ before and after the cell test at 550 ºC. **(F)** TEM images of $Al_2O_3$ before and after the cell test at 550 ºC.

### ***Effects of compacting pressure and $Al_2O_3$ layer thickness on cell performance***

The compacting-pressure effect of fabricating a cell disk on its performance was examined. As shown in **Fig. 5A**, the peak power density decreased from 1036 to 207 mW $cm^{-2}$ with increasing the fabrication compacting-pressure from 100 to 1500 MPa. This can be attributed to the decrease in proton conductivity (**Fig. 5B**), which is due to the diminished porosity (**fig. S14**). However, a relatively low compacting-pressure of 50 MPa also exhibited a lower peak power density (943 mW $cm^{-2}$) than that with 100 MPa, indicating that the pressure of 50 MPa is insufficient to compact $Al_2O_3$ particles close enough for the formation of the superstructured water at the interface between the particles, which is supported by its lower proton conductivity compared with that using 100 MPa (**Fig. 5B**). Therefore, the optimized compacting pressure for the cell fabrication is 100 MPa. Furthermore, the cell performance and the $Al_2O_3$ proton conductivity are also dependent on the thickness of the $Al_2O_3$ layer with 0.3 mm as the ideal one (Fig. 5C and 5D). Different from the conventional PCFCs that require more expensive proton conductors and complicated fabrication procedures, the WSSFC using water-superstructured porous $Al_2O_3$ layer as an unusual proton conductor can be easily fabricated with the powders of γ-$Al_2O_3$ and the electrode material (LNCA) via one-step pressing process at 100 MPa. This can greatly reduce the cell cost.

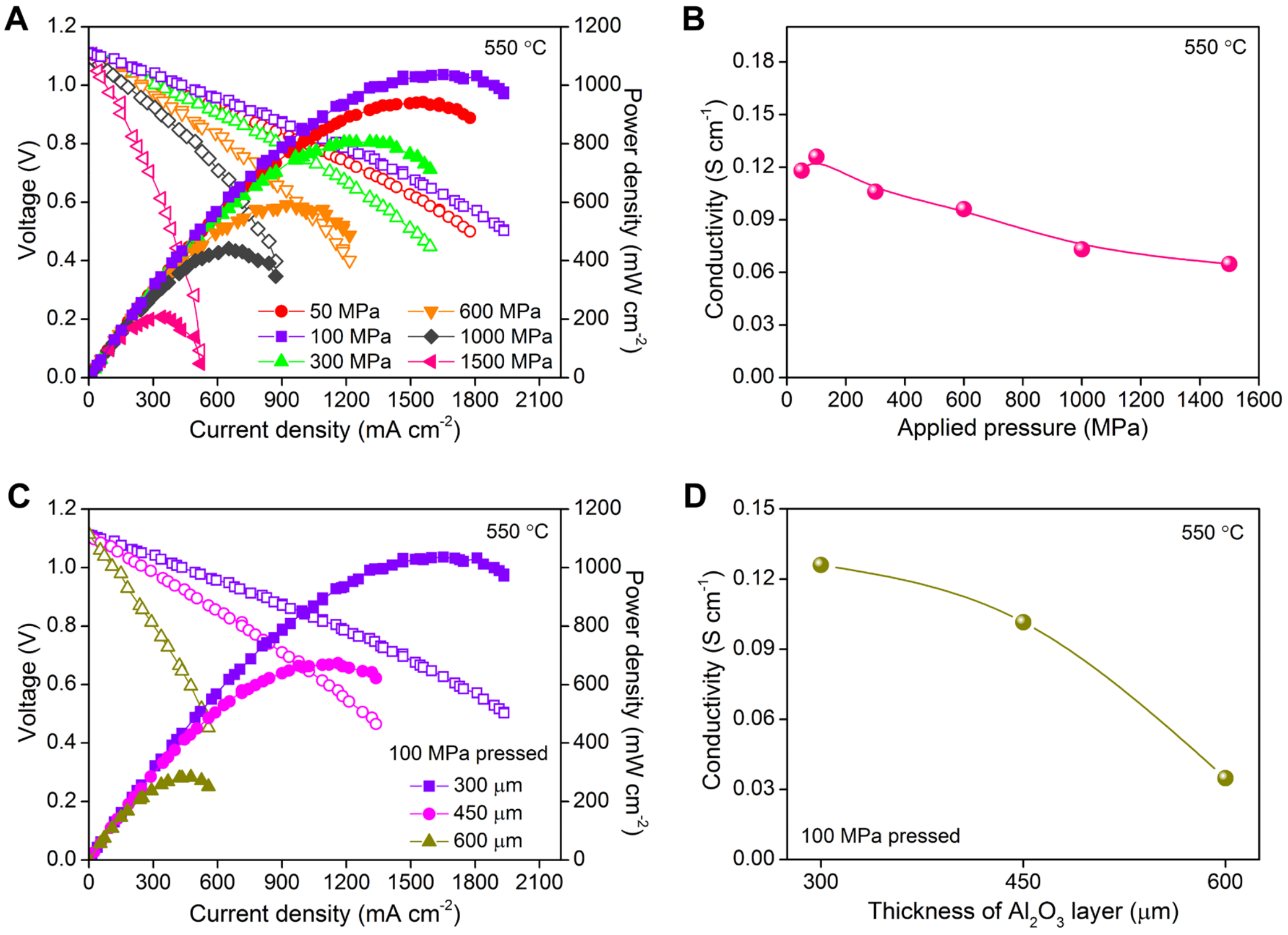

**Fig. 5. Effect of WSSFC fabrication compacting pressure on its performance. (A)** I-V and I-P curves of WSSFCs (fabricated with various compacting pressures from 50 to 1500 MPa) with $H_2$ fuel at 550 °C. **(B)** Proton conductivity of water-superstructured porous $Al_2O_3$ layer vs. its fabrication compacting pressure. **(C)** I-V and I-P curves of WSSFCs with various thicknesses of the $Al_2O_3$ layer with $H_2$ fuel at 550 °C. **(D)** Proton conductivity of water-superstructured porous $Al_2O_3$ layer vs. its thickness.

## Conclusion

This work demonstrated a new type of fuel cell—the water-superstructured solid fuel cell (WSSFC), in which the in-situ generated water-superstructured porous $Al_2O_3$ layer plays as an unprecedented superconductive electrolyte for fast proton transfer. The WSSFC exhibited very high power densities of 1036 mW $cm^{-2}$ at 550 ºC and 454 mW $cm^{-2}$ at 450 ºC as well as high OCV above 1.1 V when operated with $H_2$ as the fuel. It would be promising for commercial applications due to its excellent performance, easy fabrication, and low cost. Furthermore, the in-situ formation of superstructured-water in porous metal oxide films would be a general strategy to develop protonic superconductors and solid fuel cells.

**Acknowledgments:**

**Funding:**

U.S. National Science Foundation grant CMMI-1661699 (WZ, SF, HS, YHH)

**Author contributions:**

YHH developed the intellectual concept, designed the fuel cell structure, and provided supervisory guidance on the experiments. WZ prepared materials, fabricated fuel cells, tested cell performances, and conducted characterizations of EIS, GC, XRD, SEM, and TEM. SF conducted the measurements of in-situ FTIR and AFM as well as TG, the determination of surface areas and pore size distributions, and the quantitative analysis of XPS spectra and SEM images. HS explored electrode materials. WH, BY and GSH conducted the DFT computational work with deep analysis. KS conducted the measurements of XPS and STEM-EELS. CC and YZ conducted the X-ray computed tomography analysis of cell disks. All contributed to the data analysis. YHH wrote the manuscript with input from all other authors.

**Competing interests:**

YHH, WZ, and SF are named as inventors in US patent application on water-superstructured solid fuel cells (WSSFCs) submitted by Michigan Technological University on June 16, 2021.

**Data and materials availability:**

All data are available in the main text or the supplementary materials.

**Supplementary Materials**

Materials and Methods

Figs. S1 to S15

Tables S1 to S7

# Supplementary Materials for

## Water-Superstructured Solid Fuel Cells

Wei Zhang[1], Siyuan Fang[1], Hanrui Su[1], Wei Hao[2], Bohak Yoon[2], Gyeong S. Hwang[2], Kai Sun[3], Chung-Fu Chen[4], Yu Zhu[4], Yun Hang Hu[1]*

Correspondence to: yunhangh@mtu.edu

**This PDF file includes:**

## Materials and Methods

Materials and fuel cell disk fabrication

The commercial ultra-pure grade γ-$Al_2O_3$ powder (99.99%, Inframat Advanced Materials) was thermally treated at 900 °C in air for 10 h. The $LiNi_{0.8}Co_{0.15}Al_{0.05}O_2$ (LNCA) powder (NEI Corporation) was used as the electrode material, which was mixed with terpineol, coated on one side of Ni-foam and dried at 80 °C for 24 h to obtain LNCA electrode pieces. Fuel cell devices were fabricated by a simple dry-pressing method. Namely, one LNCA electrode piece, the γ-$Al_2O_3$ powder, and another LNCA electrode piece were layer-by-layer put into a steel die to form a symmetric LNCA/$Al_2O_3$/LNCA configuration, and then uniaxially compacted into a cylindrical porous disk. The applied pressure for the compaction ranged from 50 to 1500 MPa. All the produced cell disks have an active area of 0.64 $cm^2$ with thickness of 1-1.5 mm. Notably, although the cell disk does not possess a typical electrolyte, the $Al_2O_3$ layer between the two electrodes will be activated by in-situ produced water during the cell operation, forming water-superstructured $Al_2O_3$ layer as a proton superconductive electrolyte (see next section).

Electrochemical performance tests of water-superstructured fuel cells

A fuel cell disk was mounted and sealed on a custom-designed testing holder for performance tests. Ultra-high-purity hydrogen (99.999%, 60-80 ml $min^{-1}$) and ambient air (~100 ml $min^{-1}$) were used as fuel and oxidant at 1 atmosphere, respectively. Both the electrodes were in-situ reduced by hydrogen at 550 °C for 0.5 h before performance tests. During the cell operation, the in-situ produced water can be adsorbed on the surface of $Al_2O_3$ to create the water-superstructured $Al_2O_3$ layer as the proton superconductive electrolyte. I-V (current-voltage) and I-P (current-power) characteristics were measured by a programmable DC electronic load (IT8511A+, TEquipment). The gas leaking was tested by on-line gas chromatography (Hewlett Packard 5890 Series II equipped with an Alltech Porapak Q column and a thermal conductivity detector).

The electrical conductivity of the electrolyte was evaluated by electrochemical impedance spectra (EIS) in $H_2$/air atmospheres. The data were recorded by an electrochemical workstation (CHI760E, CH Instruments) under open-circuit conditions from $10^6$ to 0.1 Hz, with an amplitude of 10 mV at temperatures ranged from 450 to 550 °C. The measured data were fitted by the ZSimpWin 3.60 software (Ametek Scientific Instruments) to identify the electrolyte and the electrode process. The activation energy ($E_a$) was calculated with the following equation:

$$\sigma = \frac{A}{T}\exp\left(\frac{-E_a}{kT}\right) \quad (1)$$

where σ is the measured conductivity, T the absolute temperature, k the Boltzmann constant, and A the pre-exponential factor.

Basic characterizations of materials and devices

Phase crystal structure of powder samples was examined by the Scintag XDS2000 X-Ray powder diffractometer (XRD). Morphological and component analysis were conducted by the Hitachi S-4700 field emission scanning electron microscope (FE-SEM) and the FEI 200 kV Titan Themis scanning transmission electron microscope (STEM). The porous structure of the cell disk was evaluated by X-ray computed tomography (CT) technique with a Skyscan 1172 μCT instrument (under 59 kV, 169 μA and 10 W with a 0.5 mm Al filter) and its porosity was calculated with the CTan software. The $N_2$ adsorption-desorption isotherm, total BET surface area, and pore size distribution were measured on a Micromeritics ASAP 2000 adsorption instrument at 77 K. Before measurement, the samples were degassed in vacuum at 120 °C for 8 h. Furthermore, the three-dimensional morphology and size of $Al_2O_3$ particles were acquired on an Asylum Research

MFP-3D Origin+ atomic force microscope (AFM) working in the tapping mode with the 1-nm silicon probe. One hundred $Al_2O_3$ particles were imaged for statistical analysis of the external surface area. The chemical states of $Al_2O_3$ and LNCA were characterized by a PHI 5800 X-ray photoelectron spectrometer (XPS) with Mg source anode. All binding energies were calibrated with respect to the C1s line of adventitious carbon at 284.6 eV. STEM-EELS (scanning transmission electron microscopy-electron energy loss spectroscopy) images and mappings were obtained with A JEOL-JEM 3100R05 cold-FEG TEM equipped with both probe and imaging aberration correctors. The microscope was operated at 300 keV in STEM mode. The lens setting that defines a probe size smaller than 1 Å was used. EELS spectra were collected using a K2 camera using STEM EELS SI mode.

Thermogravimetric analysis

The thermogravimetric analysis of $Al_2O_3$ pellets was carried out using the Mettler Toledo TGA/SDTA851e system. Before the measurement, $Al_2O_3$ pellet (~40 mg) was pretreated at 400 ºC for 3 h. Afterwards, $Al_2O_3$ was subjected to 10 ml $min^{-1}$ water-saturated Ar gas (i.e., about 3% $H_2O$/Ar) flow with the temperature rise from 25 to 550 ºC at 5 ºC $min^{-1}$. The temperature was held at 450, 500, and 550 ºC for 2 h respectively for reaching the adsorption-desorption equilibrium. Then, 10 ml $min^{-1}$ dry Ar gas flow was introduced to the analyzer at 550 ºC for 2 h, followed by 800 ºC for another 2 h to get the absolute weight of $Al_2O_3$.

In-situ Fourier-transform infrared spectroscopy

Fourier-transform infrared spectroscopy (FTIR) was carried out on a Shimadzu IRAffinity-1 spectrometer with an in-situ diffuse reflection cell (DiffusIR, PIKE Technologies) which is equipped with a ZnSe window and can be heated up to 800 ºC. To investigate the water adsorption ability of $Al_2O_3$ at elevated temperatures, $Al_2O_3$ in a porous ceramic cup was subjected to 10 ml $min^{-1}$ water-saturated Ar gas (i.e., about 3% $H_2O$/Ar) flow at room temperature, followed by heating with rate of 2 ºC $min^{-1}$ to selected temperatures (25, 100, 200, 300, 400, 500, and 550 ºC), each of which was kept for 1 hour to ensure the water adsorption-desorption equilibrium before recording spectra. The FTIR spectra were collected in the region from 1000 to 4000 $cm^{-1}$. Afterwards, $Al_2O_3$ was heated from 550 to 800 ºC at 2 ºC $min^{-1}$ in vacuum and maintained at 800 ºC for 12 h to obtain the reference spectrum without any water. The amount of water adsorbed on $Al_2O_3$ at each selected temperature was determined from the in-situ FTIR spectrum in the range from 1500 to 4000 $cm^{-1}$.

Furthermore, to elucidate the role of the hydroxyl group of $Al_2O_3$ in proton transfer, isotope exchange experiment with the use of deuterated water ($D_2O$, 99.9 atom% D, Sigma-Aldrich) was carried out. Namely, after pressing $Al_2O_3$ into the ceramic cup, 3 droplets of water ($H_2O$) were dripped onto the $Al_2O_3$. Then, after 20-min stabilization to ensure all the $Al_2O_3$ surface to be covered by $H_2O$ and the sufficient formation of surface hydroxyl groups, 3 droplets of $D_2O$ were dripped onto the $H_2O$-covered $Al_2O_3$, followed by heating from 25 to 100, 200, 300, 400, 500, and 550 ºC in vacuum at 2 ºC $min^{-1}$ and held at each of these temperatures for 1 h before collecting the spectra. Afterwards, the temperature was elevated to 800 ºC to acquire the spectrum without any $H_2O$ and $D_2O$.

Inductively coupled plasma-optical emission spectroscopy (ICP-OES)

The accurate contents of Li in 5 parts of the used cell disk (the anode, the anode-$Al_2O_3$ interface, the middle $Al_2O_3$ layer, the cathode-$Al_2O_3$ interface, and the cathode) were determined by a PerkinElmer Optima 7000DV inductively coupled plasma-optical emission spectrometer (ICP-OES). To ensure the dissolution of Li, all the samples were previously treated by aqua regia for 3 days.

Calculation methods

The proton transfer calculations were performed on the basis of density functional theory (DFT) within the generalized gradient approximation (GGA), as implemented in the Vienna ab initio simulation package (VASP) (*S1*). The projector augmented wave (PAW) method with a plane wave basis set was employed to describe the interaction between the core and valence electrons. The valence configurations employed to construct the ionic pseudopotentials are $3s^2 3p^1$ for Al, $2s^2 2p^4$ for O and $1s^1$ for H. An energy cutoff of 450 eV was applied for the plane wave expansion of the electronic eigenfunctions. For the Brillouin zone integration, we used a (3×3×3) Monkhorst-Pack mesh of k points to determine the optimal geometries and total energies of the $\gamma$-$Al_2O_3$ bulk and (1×1×1) Monkhorst-Pack mesh of k points for $\gamma$-$Al_2O_3$ (100) surface with a monolayer (ML) of $H_2O$. Reaction pathways and barriers were determined using the climbing image nudged elastic band (ci-NEB) method with five intermediate images for each elementary step.

The unit cell of pristine $\gamma$-$Al_2O_3$ bulk was constructed based on the data of work done by Digne et al (*S2*). A 2×1×1 slab structure with a 15 Å vacuum was set up to simulate the (100) surface of $\gamma$-$Al_2O_3$. Totally 10 $H_2O$ molecules, which formed a ML of $H_2O$, were added on the top surface of $\gamma$-$Al_2O_3$. An additional proton was inserted in the ML of $H_2O$ and optimized to the stable state. For all slab structures, the atoms were fully relaxed with the conjugate gradient method until residual forces on all of the constituent atoms became smaller than $2\times10^{-2}$ eV/Å. The similar model was used for evaluation of proton transfer on the (100) surface of $\gamma$-$Al_2O_3$ with an adsorbed $H_2O$ molecule, but the water ML consisting of 10 $H_2O$ molecules was replaced by single $H_2O$ molecule.

**Structure and property characterization of γ-$Al_2O_3$**

DFT calculations for path and barrier of proton transfer

**Fig. S1.**

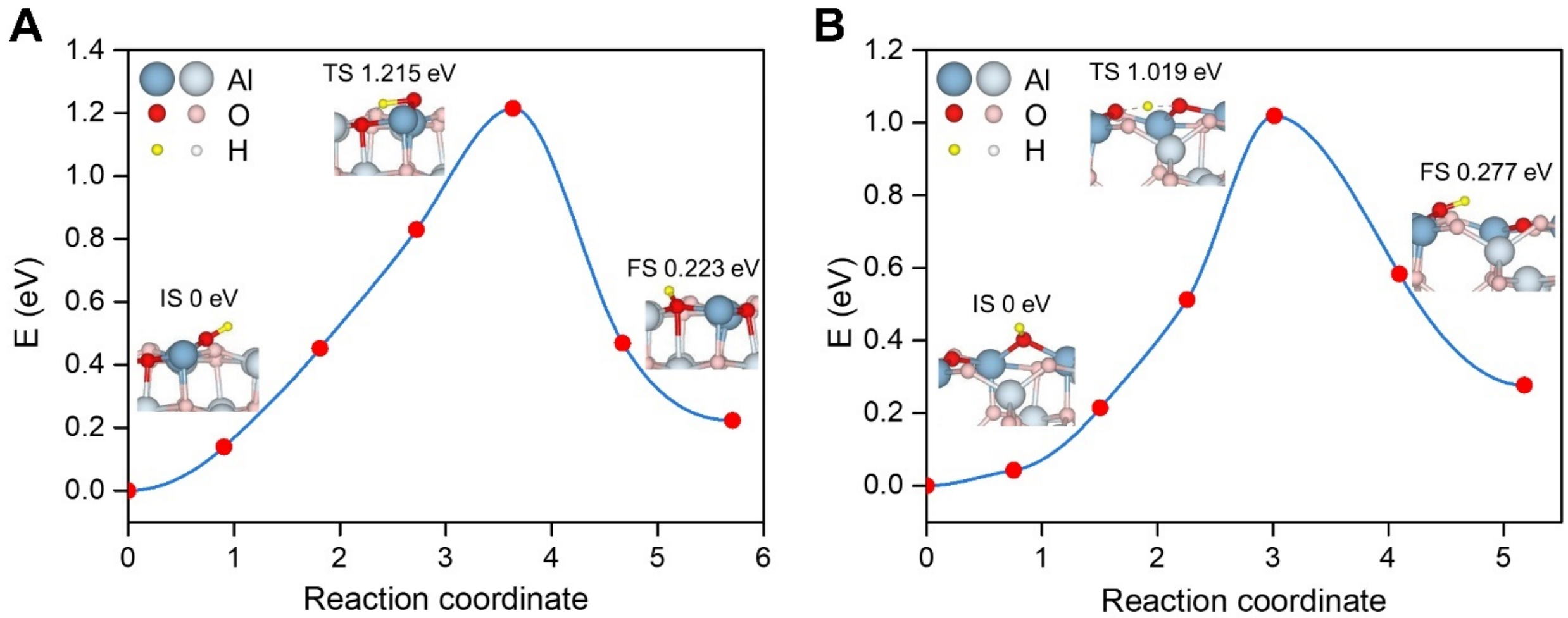

**Fig. S1.** Reaction path and barrier of proton transfer on the (100) surface of γ-$Al_2O_3$ for two most possible cases **(A)** and **(B)**. The insertion figures show the initial, transition and final states of the proton transfer.

**Fig. S2.**

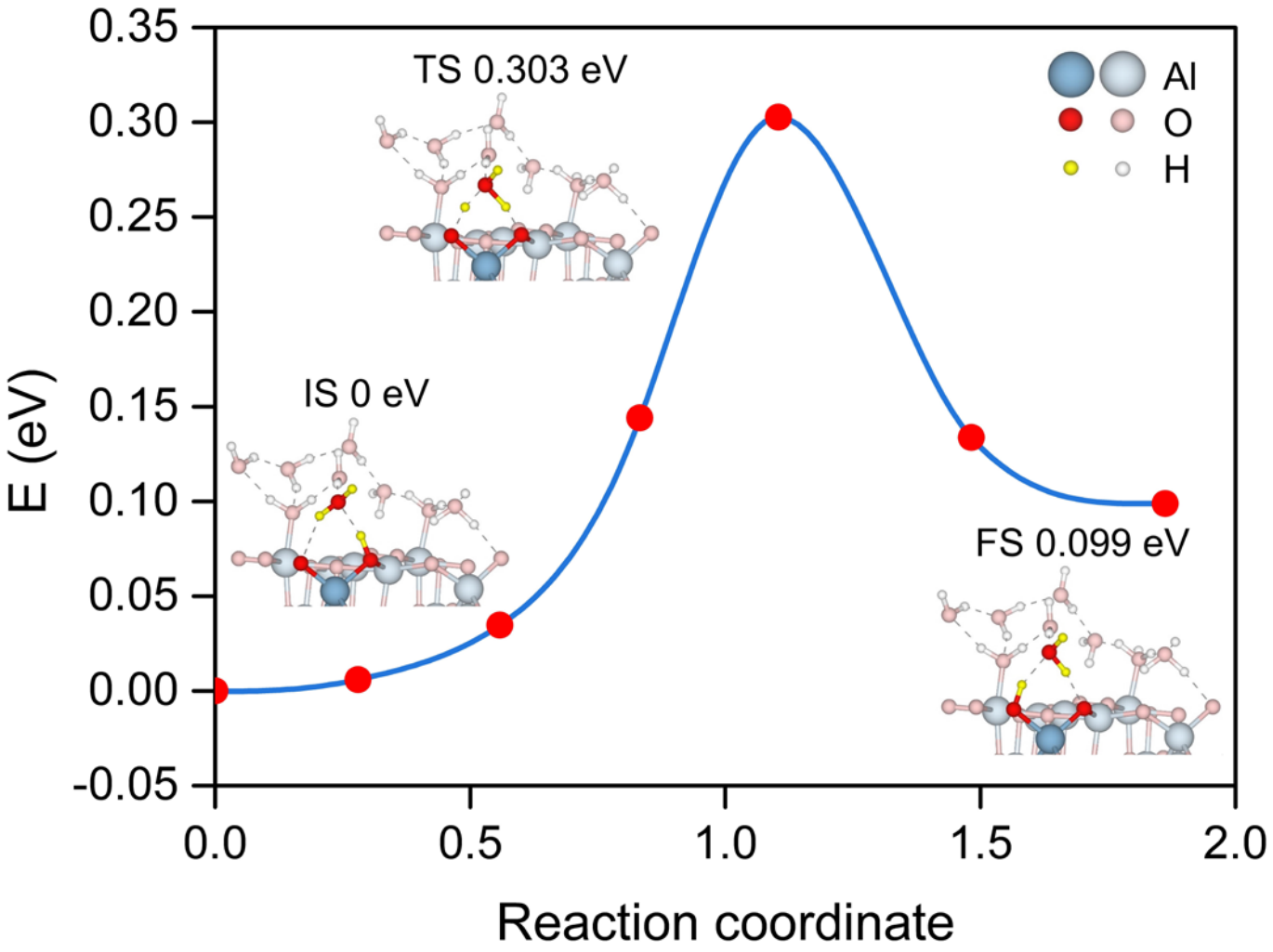

**Fig. S2.** Water-mediated proton transfer on $Al_2O_3$ surface with a water monolayer demonstrated by DFT calculation.

AFM image of $\gamma$-$Al_2O_3$ particles

**Fig. S3.**

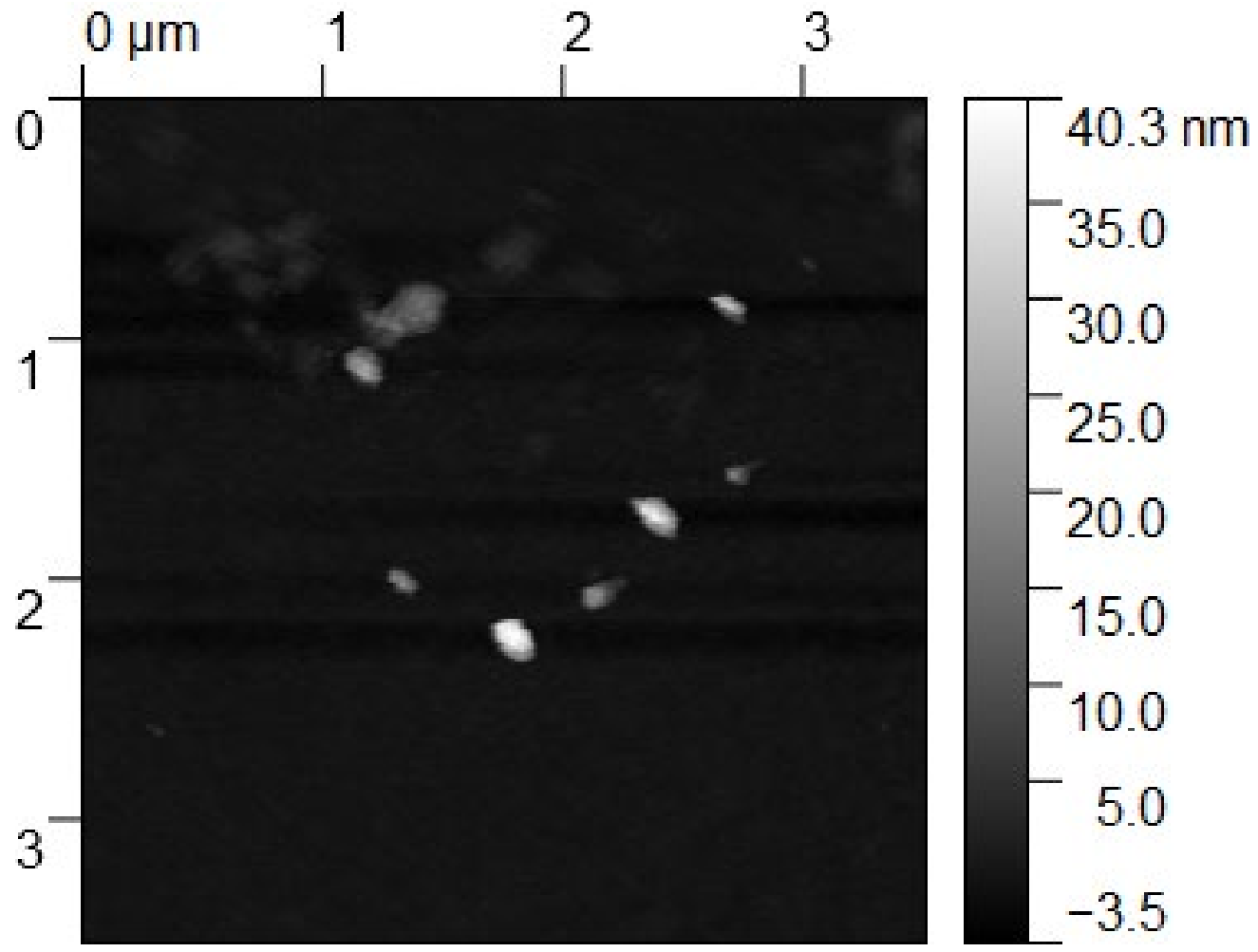

**Fig. S3.** AFM image of $Al_2O_3$ particles.

**Fig. S4.**

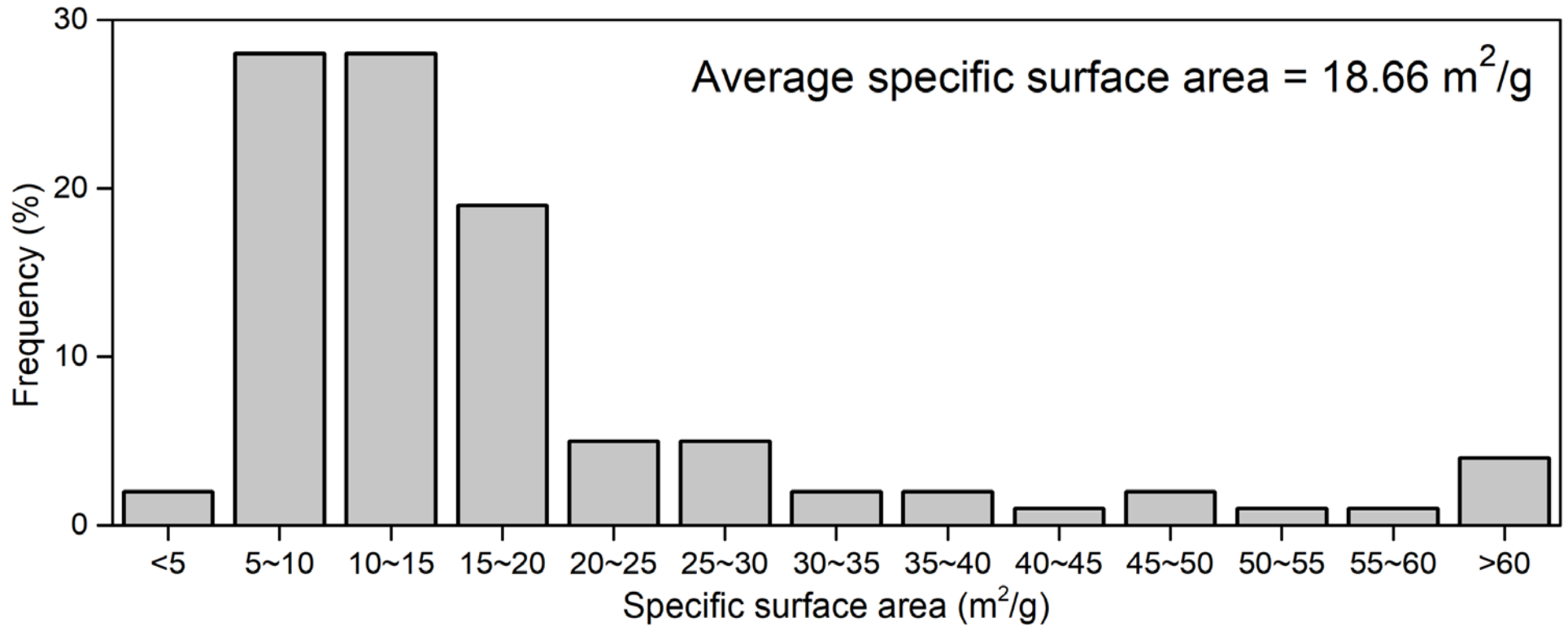

**Fig. S4.** Specific external surface area distribution of γ-$Al_2O_3$ powder from AFM images.

BET surface area and pore distribution of γ-$Al_2O_3$ powder

**Fig. S5.**

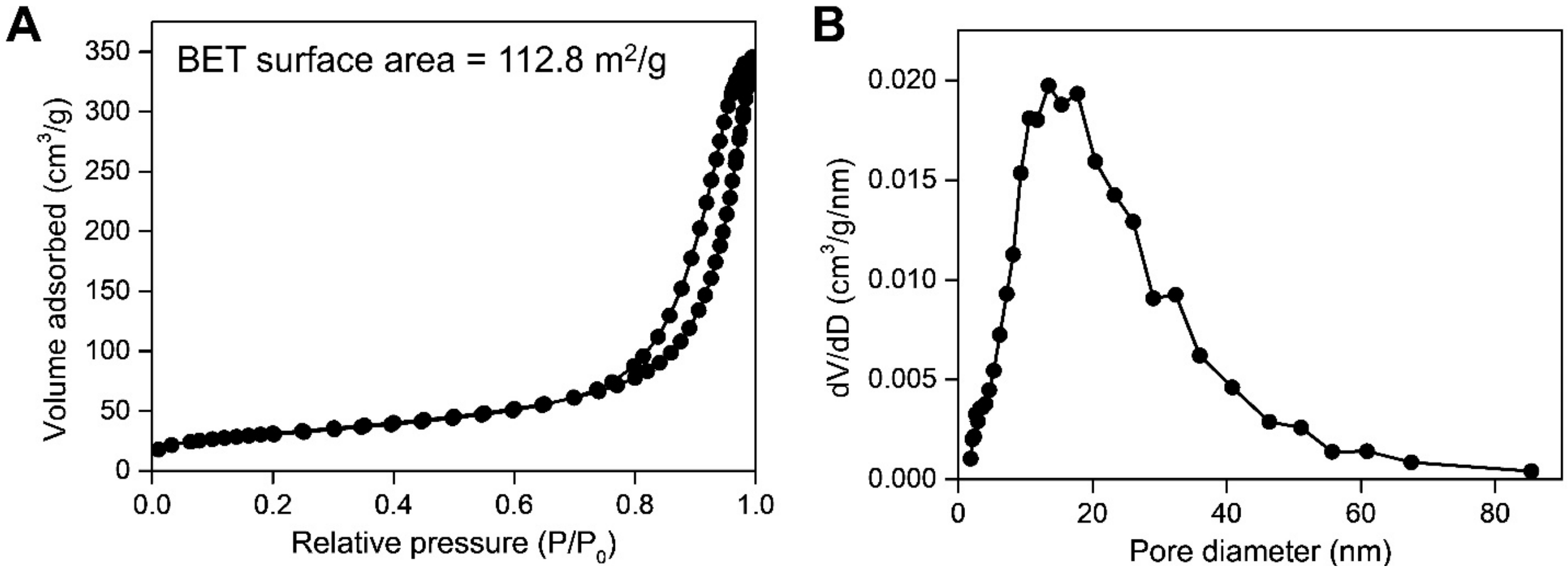

**Fig. S5.** Total surface area and pore structure of $Al_2O_3$ powder. **(A)** $N_2$ adsorption-desorption isotherm curves at liquid nitrogen temperature (-196 °C) and BET surface area. **(B)** BJH pore size distribution.

Thermogravimetric (TG) evaluation of water on γ-$Al_2O_3$ pellets

**Table S1.**

**Table S1.** Water coverage on $Al_2O_3$ pellet (fabricated by pressing $Al_2O_3$ powder) under 3% $H_2O$/argon flow from TG measurements.

| **$Al_2O_3$ pellet fabrication** | **Water adsorption on $Al_2O_3$ pellet** | | |
|---|---|---|---|
| **Pressing pressure (MPa)** | **Temperature ($^oC$)** | **Adsorbed water (mg $H_2O$/g $Al_2O_3$)** | **Water coverage area ($m^2$ $H_2O$/g $Al_2O_3$)** |
| 50 | 450 | 7.63 | 5.10 |
| 50 | 500 | 5.99 | 4.01 |
| 50 | 550 | 4.36 | 2.92 |
| 100 | 450 | 6.48 | 4.33 |
| 100 | 500 | 4.86 | 3.25 |
| 100 | 550 | 4.05 | 2.71 |
| 1500 | 450 | 5.48 | 3.67 |
| 1500 | 500 | 3.99 | 2.67 |
| 1500 | 550 | 2.99 | 2.00 |

**Table S2.**

**Table S2.** Water coverage on $Al_2O_3$ pellet (fabricated by pressing $Al_2O_3$ powder at 100 MPa) from TG measurements at 550 °C.

| Water concentration in argon flow | Adsorbed water (mg $H_2O$/g $Al_2O_3$) | Water coverage area ($m^2$ $H_2O$/g $Al_2O_3$) |
|---|---|---|
| 3% | 4.05 | 2.71 |
| 9% | 9.01 | 6.02 |

**Characterization of WSSFC fabricated with 100 MPa compacting pressure**

Porosities of electrodes and $Al_2O_3$ layer from 3D X-ray CT and SEM images

**Table S3.**

**Table S3.** Porosity of electrodes and $Al_2O_3$ layer.

| Material | Porosity from 3D X-ray CT | Porosity from SEM |
|---|---|---|
| Top electrode | 27.3% | 26.1 % |
| $Al_2O_3$ layer | 26.3% | 25.4% |
| Bottom electrode | 29.8% | 28.3% |

BET surface area and pore distribution of γ-$Al_2O_3$ pellet

**Fig. S6.**

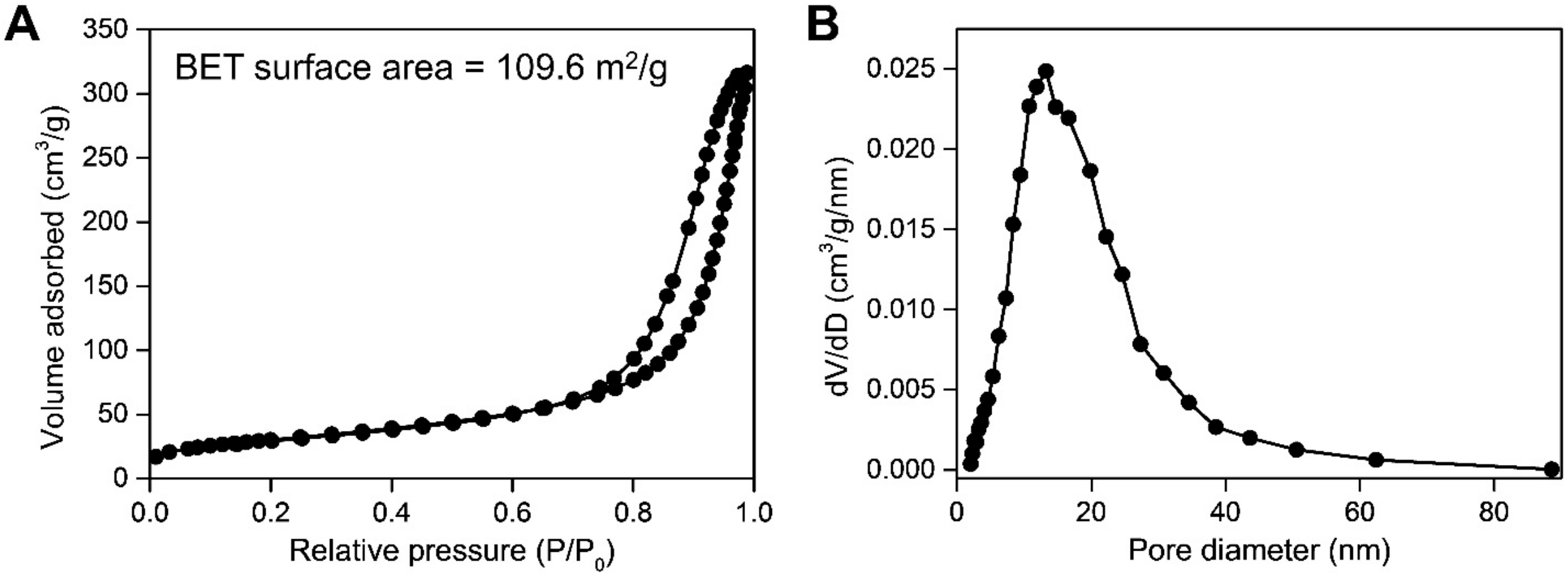

**Fig. S6.** Total surface area and pore structure of $Al_2O_3$ pellet. **(A)** $N_2$ adsorption-desorption isotherm curves at liquid nitrogen temperature (-196 °C) and BET surface area. **(B)** BJH pore size distribution.

Gas leakage test for a cell disk without a drilled hole

To confirm the diffusion of gas through the $Al_2O_3$ layer between the anode and the cathode, $H_2$ and air were introduced into the anode and cathode sides, respectively. As shown in **fig. S7A**, $H_2$ fed to the anode was detected by Gas Chromatography (GC) in the cathode side without cell operation at room temperature, revealing that $H_2$ transferred through the $Al_2O_3$ layer. In contrast, when the cell was being operated at 550 °C, $H_2$ peak almost disappeared in the GC spectrum (**fig. S7B**), indicating the negligible gas transfer through the $Al_2O_3$ layer during the cell operation, which is consistent with the obtained high open circuit voltage (OCV) of 1.11 V.

**Fig. S7.**

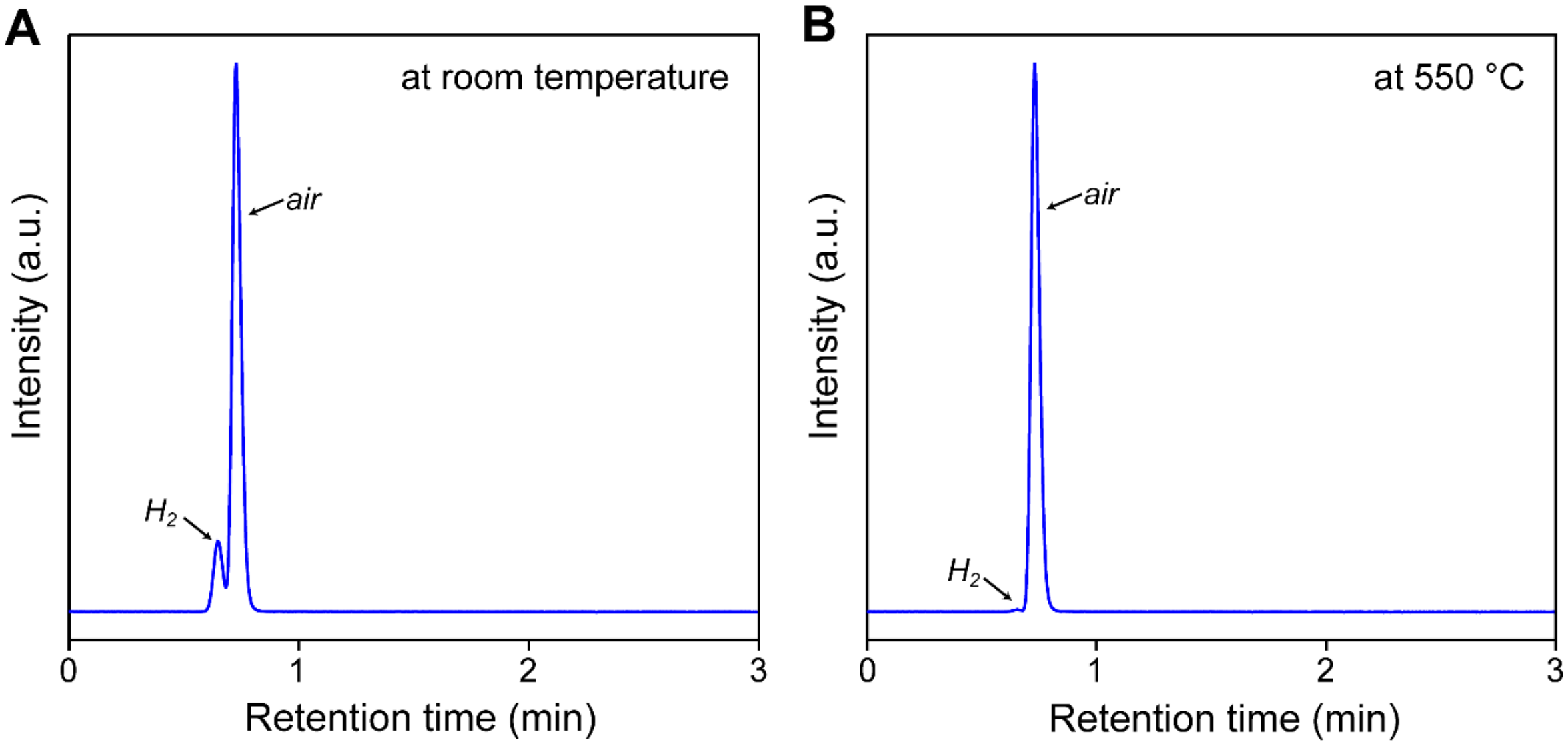

**Fig. S7.** GC spectra of gases in the cathode side of the LNCA/γ-$Al_2O_3$/LNCA cell, in which $H_2$ and air were respectively introduced into its anode and cathode **(A)** at room temperature (cell OCV: 0 V) and **(B)** at 550 °C (cell OCV: 1.11 V).

Effect of gas leakage on OCV for a cell disk with a drilled hole

To evaluate the effect of gas leakage on OCV, we on-purpose drilled a hole (1 mm diameter) in the cell disk, thus the large $H_2$ peaks were observed in the GC spectra for both cases with and without operation, confirming the gas diffusion through the $Al_2O_3$ layer even during the operation at 550 °C (**fig. S8**) and thus leading to a negligible OCV of 0.02 V. This clearly demonstrated that the gas leakage in a cell can tremendously decrease its OCV.

**Fig. S8.**

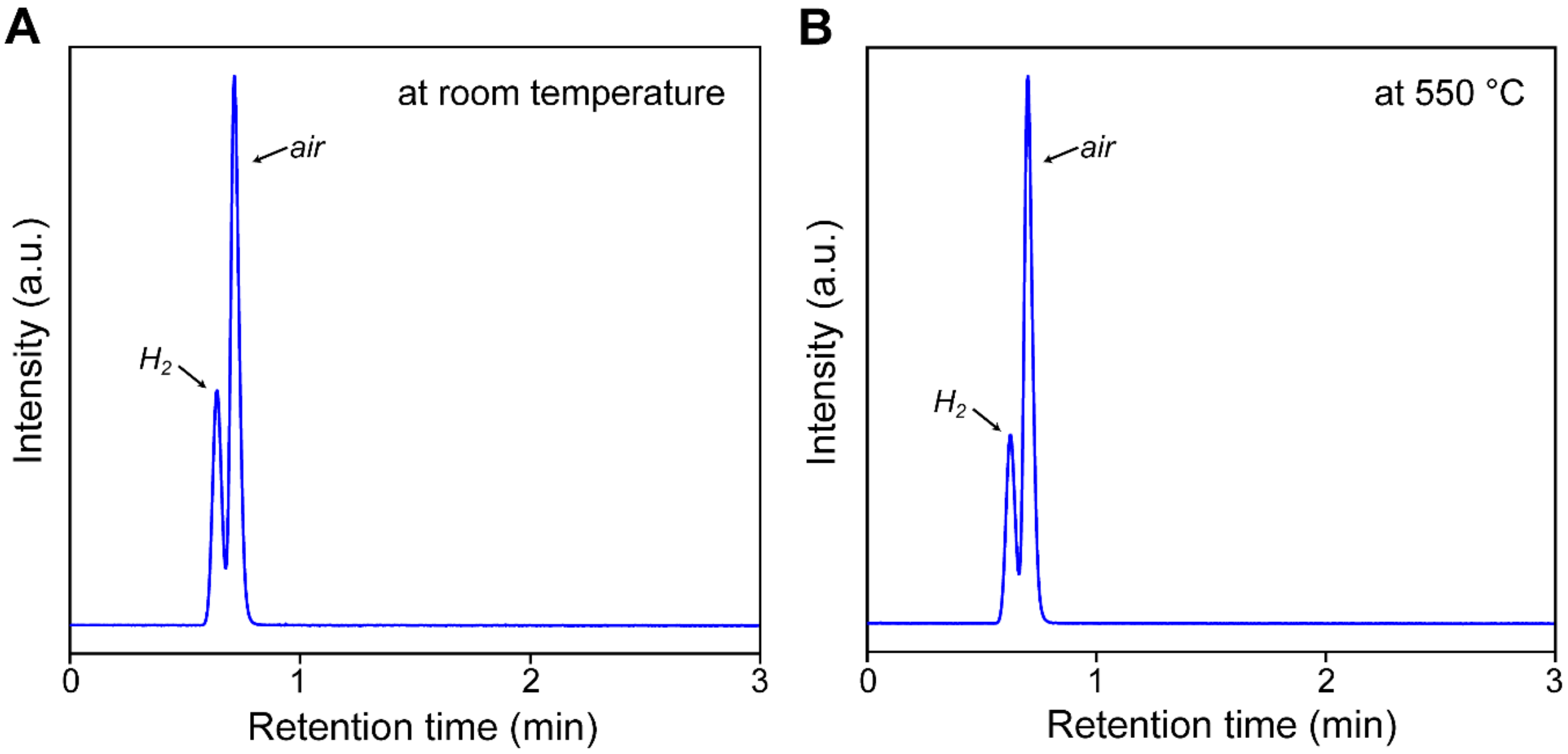

**Fig. S8.** GC spectra of gases in the cathode side of the LNCA/γ-$Al_2O_3$/LNCA cell with a drilled hole (1 mm hole diameter) through the cell disk, in which $H_2$ and air were respectively introduced into its anode and cathode **(A)** at room temperature (cell OCV: 0 V) and **(B)** at 550 °C (cell OCV: 0.02 V).

Exploration of electrodes

**Table S4.**

**Table S4.** Tests of anodes and cathodes for the WSSFC fed with $H_2$ fuel for anode and air for cathode at 550 °C.

| Anode | Electrolyte | Cathode | Peak Power Density ($mW/cm^2$) | OCV (V) |
|---|---|---|---|---|
| NiO | $Al_2O_3$ | LSCF | 0 | 0 |
| Ni-SDC | $Al_2O_3$ | LSCF | 0 | 0 |
| Ni-BZCYYb | $Al_2O_3$ | LSCF | 0 | 0 |
| LNCA | $Al_2O_3$ | LSCF | 0 | 0.34 |
| NiO | $Al_2O_3$ | BCFZY | 0 | 0 |
| Ni-SDC | $Al_2O_3$ | BCFZY | 0 | 0 |
| Ni-BZCYYb | $Al_2O_3$ | BCFZY | 0 | 0 |
| LNCA | $Al_2O_3$ | BCFZY | 0 | 0.76 |
| NiO | $Al_2O_3$ | NiO | 0 | 0 |
| Ni-SDC | $Al_2O_3$ | NiO | 0 | 0 |
| Ni-BZCYYb | $Al_2O_3$ | NiO | 0 | 0 |
| LNCA | $Al_2O_3$ | NiO | 0 | 0.41 |
| NiO | $Al_2O_3$ | LNCA | 50 | 0.78 |
| Ni-SDC | $Al_2O_3$ | LNCA | 181 | 0.74 |
| Ni-BZCYYb | $Al_2O_3$ | LNCA | 212 | 0.78 |
| LNCA | $Al_2O_3$ | LNCA | 1036 | 1.11 |

**Note:** SDC: $Ce_{0.8}Sm_{0.2}O_{1.9}$; BZCYYb: $BaZr_{0.1}Ce_{0.7}Y_{0.1}Yb_{0.1}O_{3-\delta}$; LSCF: $(La_{0.60}Sr_{0.40})_{0.95}Co_{0.20}Fe_{0.80}O_{3-\delta}$; BCFZY: $BaCo_{0.4}Fe_{0.4}Zr_{0.1}Y_{0.1}O_{3-\delta}$; LNCA: $LiNi_{0.8}Co_{0.15}Al_{0.05}O_2$.

Porosities of electrodes, $Al_2O_3$ layer, and interfaces after cell test

**Table S5.**

**Table S5.** Porosities of electrodes, $Al_2O_3$ layer, and interfaces after cell test (from SEM images).

| **Material** | **Porosity** |
|---|---|
| Anode | 28.1% |
| Anode-$Al_2O_3$ interface | 13.2% |
| $Al_2O_3$ layer | 26.3% |
| Cathode-$Al_2O_3$ interface | 15.0% |
| Cathode | 28.8% |

XPS analysis of LNCA electrodes and $Al_2O_3$ layer

**Fig. S9.**

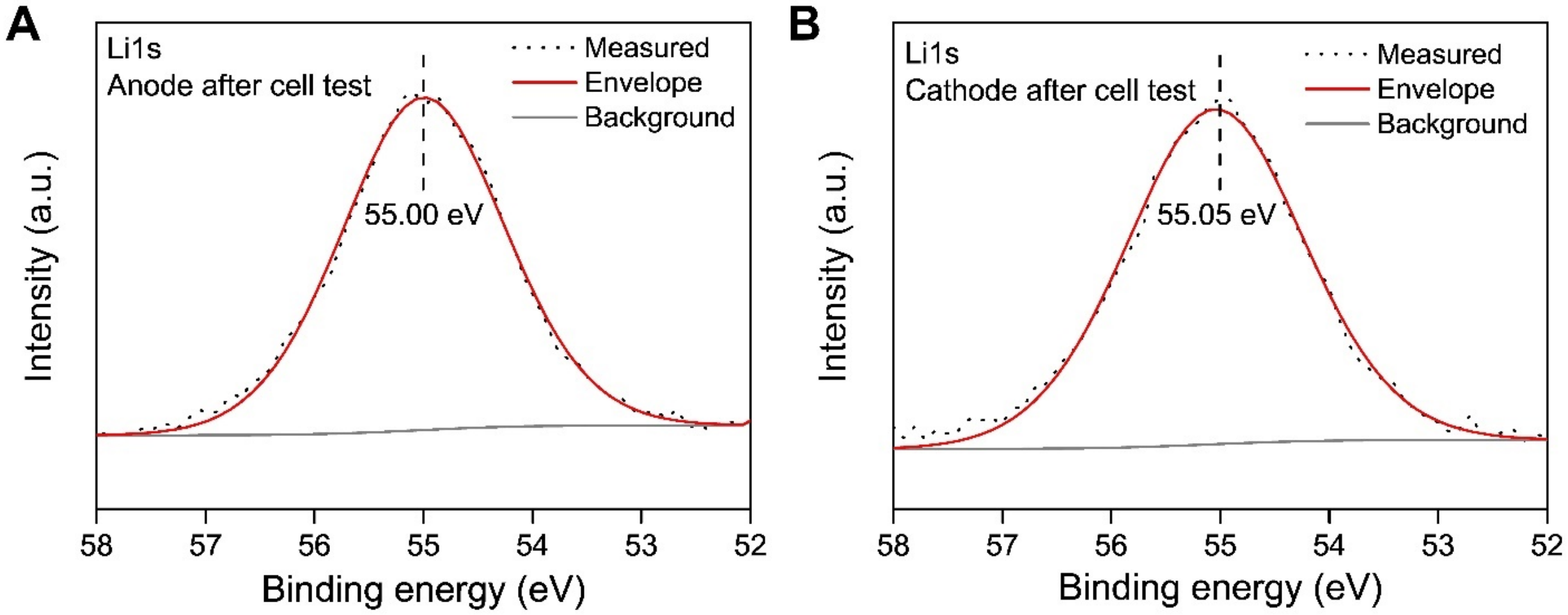

**Fig. S9.** Li 1s XPS spectra of **(A)** anode and **(B)** cathode materials after cell test.

**Fig. S10.**

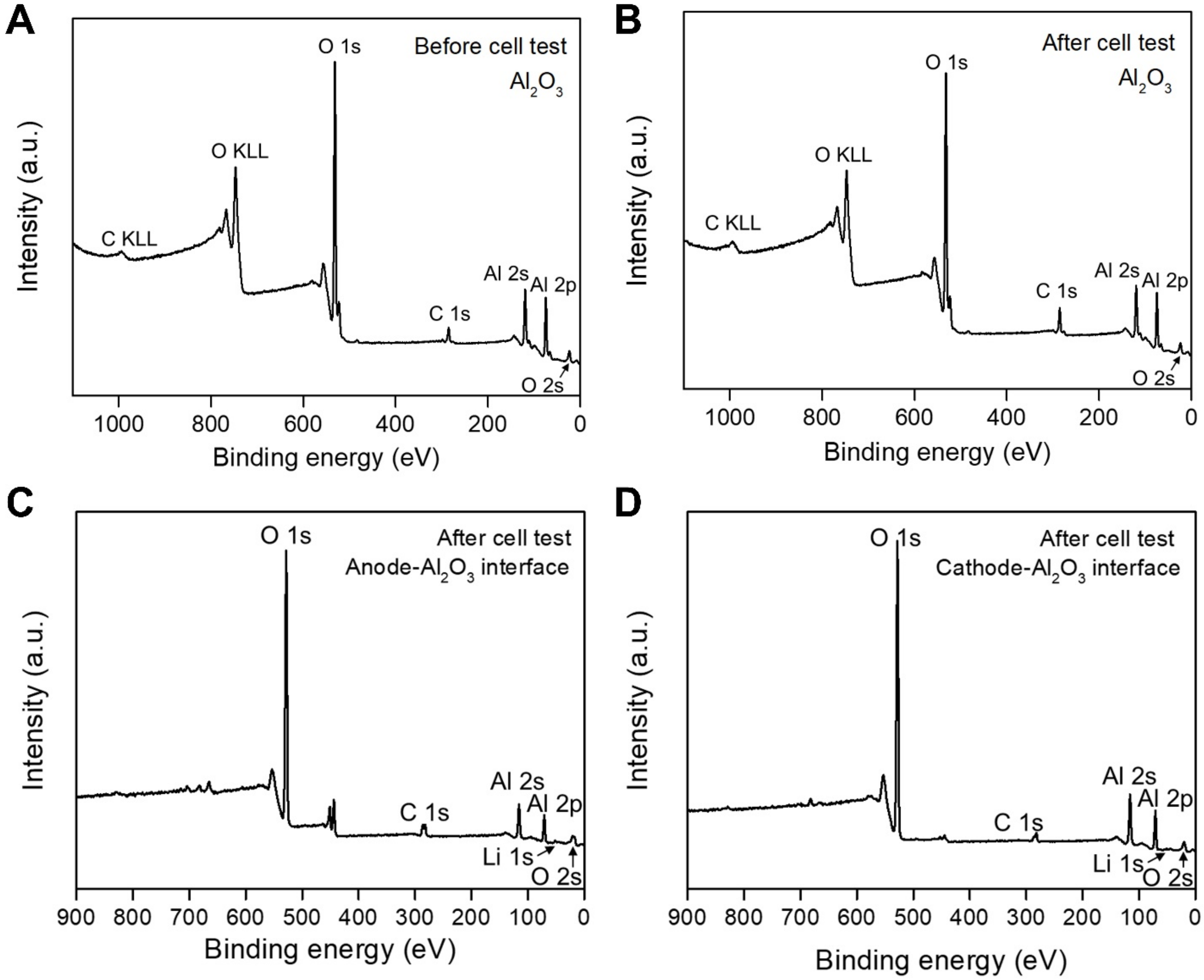

**Fig. S10.** XPS survey spectra of $Al_2O_3$ **(A)** before and **(B)** after cell test, and **(C)** anode-$Al_2O_3$ interface and **(D)** cathode-$Al_2O_3$ interface after cell test.

Effect of LiOH on cell performance

LiOH was introduced into the anode (NiO), the middle $Al_2O_3$ layer, and the cathode (NiO or $(La_{0.60}Sr_{0.40})_{0.95}Co_{0.20}Fe_{0.80}O_{3-\delta}$ named as LSCF), respectively. As shown in Table S6, when 30 wt% LiOH was added into the $Al_2O_3$ layer of the NiO/$Al_2O_3$/NiO cell, the cell operation with $H_2$ fuel for its anode and air for its cathode generated zero power density due to no current. This indicates that LiOH didn't contribute to the transfer of protons or $O^{2-}$ ions in the $Al_2O_3$ layer for the cell. Furthermore, when 25 wt% LiOH was included in the NiO anode, power density was not obtained either. In contrast, the introduction of 25 wt% LiOH into the NiO cathode generated power density of 22 mW/cm$^2$. Furthermore, when 33 wt% LiOH was added into the LSCF cathode, the cell generated power density of 89 mW/cm$^2$. These reveal that LiOH is necessary in the cathode of the cell. Different from a conventional ceramic proton fuel cell that has a sintered cathode to reduce $O_2$ to $O^{2-}$ ions and then transfer to the cathode-electrolyte interface for the reaction with protons to $H_2O$, the WSSFC possessed an un-sintered cathode that suffers the large grain boundary with high resistance for the transfer of $O^{2-}$ ions. Therefore, a LiOH liquid phase or its aqueous solution plays an important role in the cathode of WSSFC, namely, the $O^{2-}$ ions (generated on Ni/NiO active components of the cathode) combine with $H_2O$ to $OH^-$ ions which then transfer via LiOH to the interface between cathode and the $Al_2O_3$ layer to react with protons to $H_2O$. This explains why a LNCA electrode, which can decompose into not only active components of Ni/NiO but also LiOH, is needed for WSSFCs (**table S6**).

Importantly, the existence of LiOH in the cathode raises concern about whether LiOH may react with $CO_2$ in air flow to continually form $Li_2CO_3$ and thus may decrease the performance of the cathode. To clarify it, we examined the cathode by FT-IR spectra. As shown in **fig. S11**, one can see no IR carbonate peak at around 1500 cm$^{-1}$ for pristine LNCA, whereas the carbonate peak was observed for the cathode at open circuit condition and 550 $^oC$ for 2 hours. However, the peak intensity of carbonate is smaller for the cathode at the closed circuit condition than that at the open circuit condition for 2 hours, indicating that the cell operation inhibited the formation of carbonate. Furthermore, the IR peak intensity of carbonate is much smaller for 100 hours of operation than for 2 hours, indicating that the carbonate in the cathode greatly decreased with reaction time. This demonstrated that the formation of $Li_2CO_3$ in the cathode is not an issue for the cell performance.

**Table S6.**

**Table S6.** Effect of LiOH on the performance of WSSFC with $H_2$ fuel for anode and air for cathode at 550 $^oC$.

| Anode | Electrolyte | Cathode | PPD (mW/cm$^2$) | OCV (V) |
|---|---|---|---|---|
| NiO | $Al_2O_3$ | NiO | 0 | 0 |
| NiO (25wt% LiOH) | $Al_2O_3$ | NiO | 0 | 0.87 |
| NiO | $Al_2O_3$ (30wt% LiOH) | NiO | 0 | 0.46 |
| NiO | $Al_2O_3$ | NiO (25wt% LiOH) | 22 | 0.99 |
| NiO | $Al_2O_3$ | LSCF | 0 | 0 |
| NiO (25wt% LiOH) | $Al_2O_3$ | LSCF | 0 | 0.70 |
| NiO | $Al_2O_3$ (30wt% LiOH) | LSCF | 0 | 0.15 |
| NiO | $Al_2O_3$ | LSCF (33wt% LiOH) | 89 | 0.97 |

LSCF: $(La_{0.60}Sr_{0.40})_{0.95}Co_{0.20}Fe_{0.80}O_{3-\delta}$

**Fig. S11.**

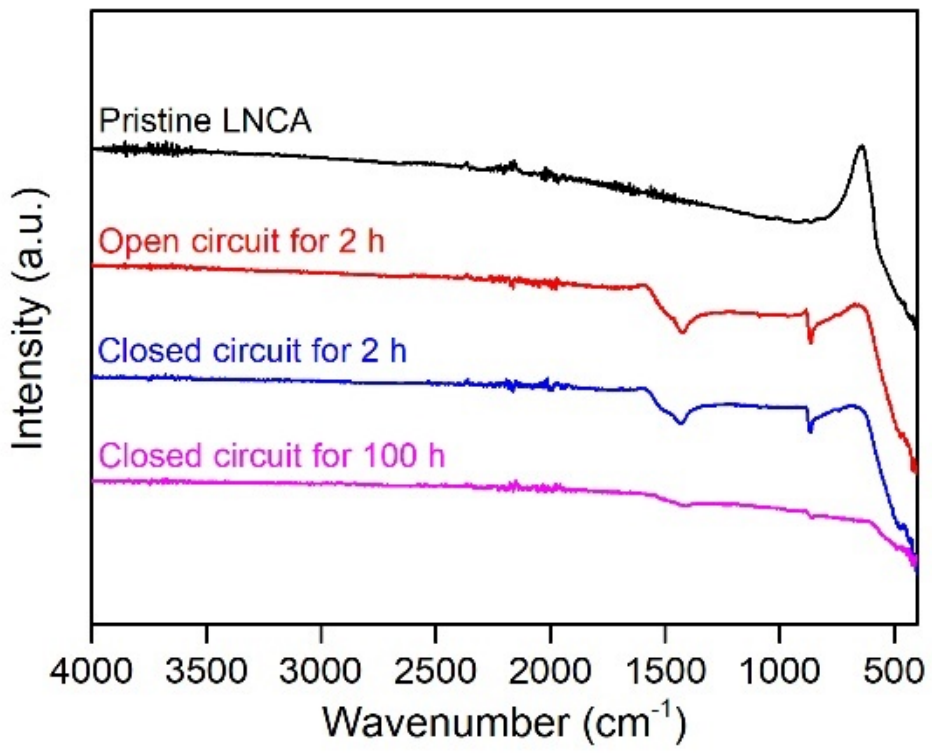

**Fig. S11.** FTIR spectra of LNCA as the air cathode used in WSSFC at various conditions (pristine LNCA without operation, after 2 hours at open circuit condition and 550 ºC, after operation at closed circuit condition (with current density of 72 mA/cm$^2$) and 550 ºC for 2 hours, and after operation at closed circuit condition (with current density of 72mA/cm$^2$) and 550 ºC for 100 hours, respectively).

Proton conductivity

The protonic conductivity of a ceramic protonic electrolyte is usually determined by electrochemical impedance spectroscopy (EIS) with inert electrodes (such as Au) under wet $N_2$ atmosphere (*S3, S4*). This is based on the following principle: the interaction between the oxygen vacancies of the ceramic proton conductor and steam can produce proton ions, leading to the protonic transportation in the conductor (*S5, S6*), First, we employed this method to determine the protonic conductivity of $Al_2O_3$ with inert Au electrodes under wet $N_2$ atmosphere, showing very high area specific resistance of $5.8\times10^7$ Ω $cm^2$ (i.e., very low conductivity of $5.1\times10^{-10}$ S $cm^{-1}$) (**fig. S12**). This is consistent with the general recognition that intrinsic $Al_2O_3$ is an ionic insulator. Unlike a ceramic protonic conductor that has rich oxygen vacancies, $Al_2O_3$ doesn't possess oxygen vacancies and thus cannot generate protons when it is in contact of steam. In contrast, the protonic conductivity of "water-superstructured $Al_2O_3$" is in-situ and dynamically created during the cell operation, namely, protons generated on the anode transfer to the cathode through the $Al_2O_3$ surface with water that was in-situ generated. For this reason, we need to determine the proton conductivity under the operation condition for the cell with the symmetric configuration of LNCA/$Al_2O_3$/LNCA (here LNCA=$LiNi_{0.8}Co_{0.15}Al_{0.05}O_2$), exhibiting ultrahigh proton conductivity of 0.13 S $cm^{-1}$ at 550 ºC (**Fig. 4A**). Obviously, such an in-situ measurement raised an important concern about whether Li ions from the decomposition of the LNCA electrode may contribute to the ionic transfer in the $Al_2O_3$ layer. However, we experimentally excluded the role of any Li-based compound in the ionic conductivity of the water-superstructured $Al_2O_3$ layer as follows: To make sure whether Li ions of the electrode entered into the $Al_2O_3$ layer, we used inductively coupled plasma (ICP) element analysis to determine Li contents in the 5 parts of the tested cell disk: LNCA anode, anode/$Al_2O_3$ interface, middle $Al_2O_3$ layer, $Al_2O_3$/cathode interface, LNCA cathode. The ICP results showed that Li was not found in the middle $Al_2O_3$ layer, though Li was detected in the anode, cathode, and two interfaces (**Fig. 3F**). Furthermore, the absence of Li in the middle $Al_2O_3$ layer was further confirmed by EELS and XPS (**Fig. 3H** and **fig. S10**).

**Fig. S12.**

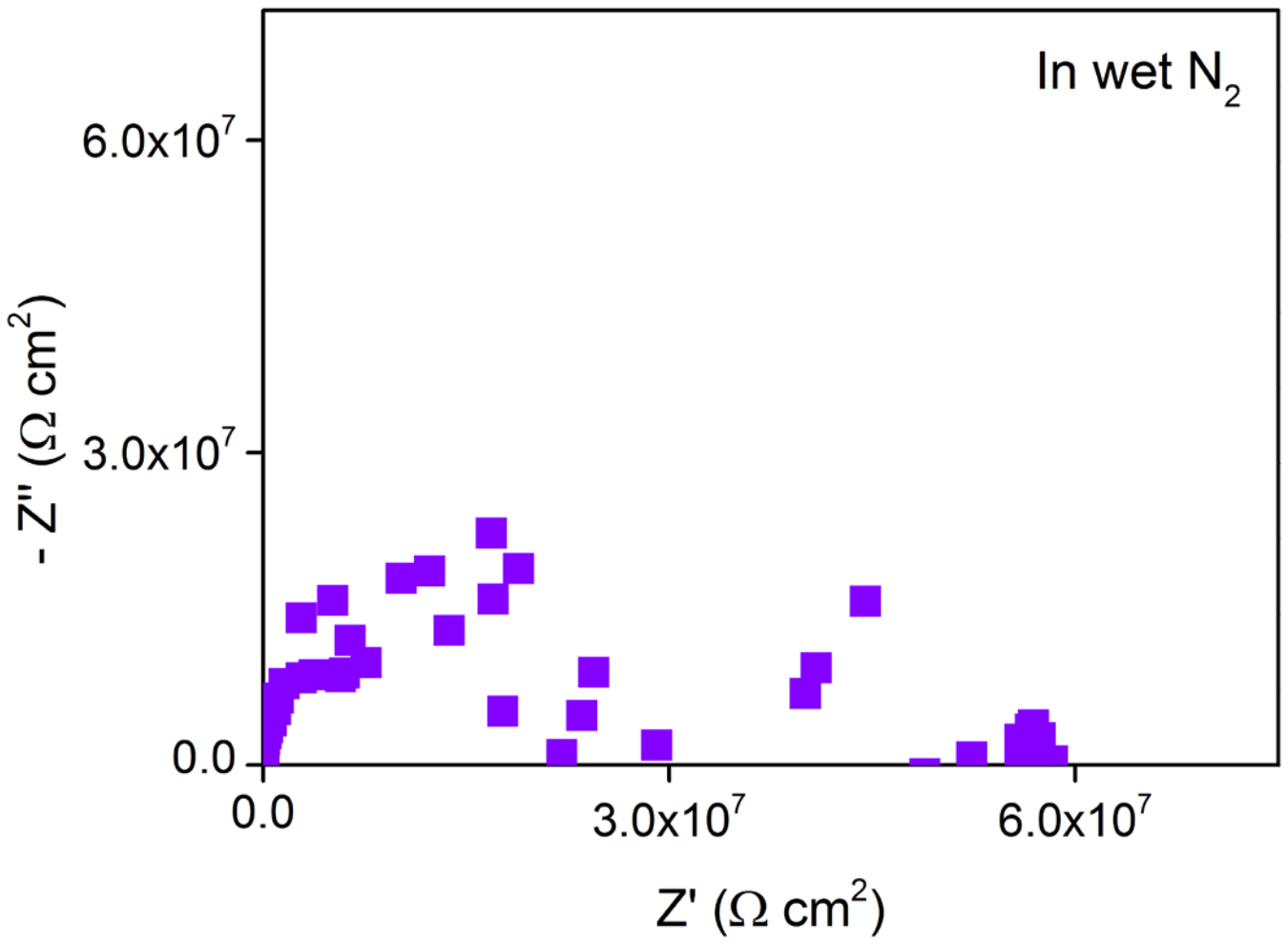

**Fig. S12.** Impedance spectrum of the Au/$Al_2O_3$/Au cell in wet $N_2$ (with 3% $H_2O$) at 550 ºC. It was measured under open-circuit conditions by changing the frequency from $10^6$ to 0.1 Hz. Note: The area specific resistance is at the level of $10^7$ Ω $cm^2$, revealing that intrinsic $Al_2O_3$ layer is an ionic insulator with very low conductivity of $5.1\times10^{-10}$ S $cm^{-1}$.

**Fig. S13.**

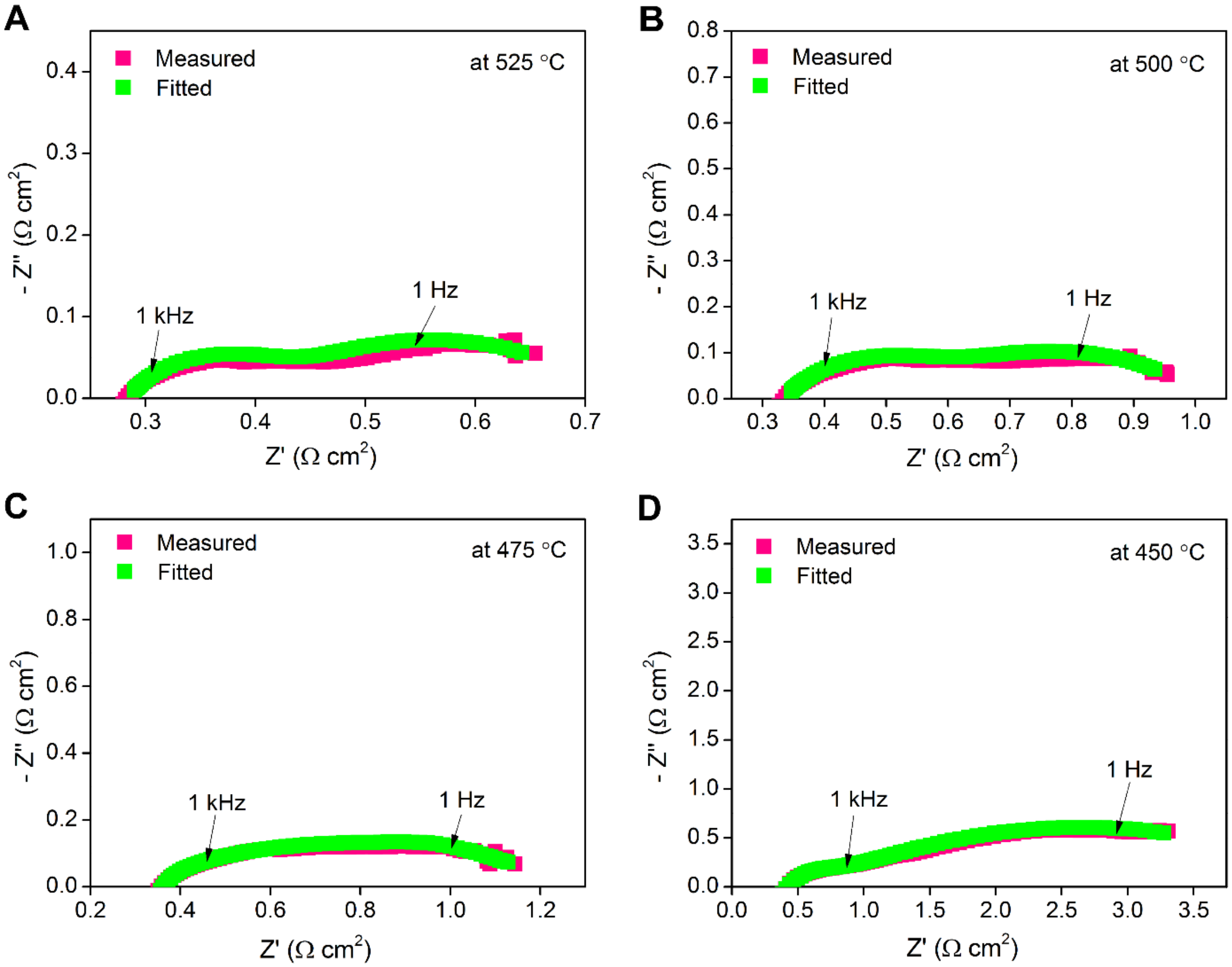

**Fig. S13**. Impedance spectra of the LNCA/$Al_2O_3$/LNCA cell under $H_2$/air open-circuit conditions at various temperatures. It was measured under open-circuit conditions by changing the frequency from $10^6$ to 0.1 Hz. They were fitted with the equivalent circuit of R0(R1||CPE1)(R2||CPE2) (shown by the inset in Fig. 4A).

**Table S7.**

**Table S7.** The area-specific ohmic resistances ($ASR_{ohm}$) obtained from the fitted impedance spectra and corresponding proton conductivities of the water-superstructured $Al_2O_3$ layer at different temperatures.

| Temperature (°C) | 450 | 475 | 500 | 525 | 550 |
|---|---|---|---|---|---|
| $ASR_{ohm}$ ($\Omega$ $cm^2$) | 0.42 | 0.36 | 0.33 | 0.28 | 0.24 |
| Proton conductivity (S $cm^{-1}$) | 0.07 | 0.08 | 0.09 | 0.11 | 0.13 |

Effect of compacting pressure on porosity

**Fig. S14.**

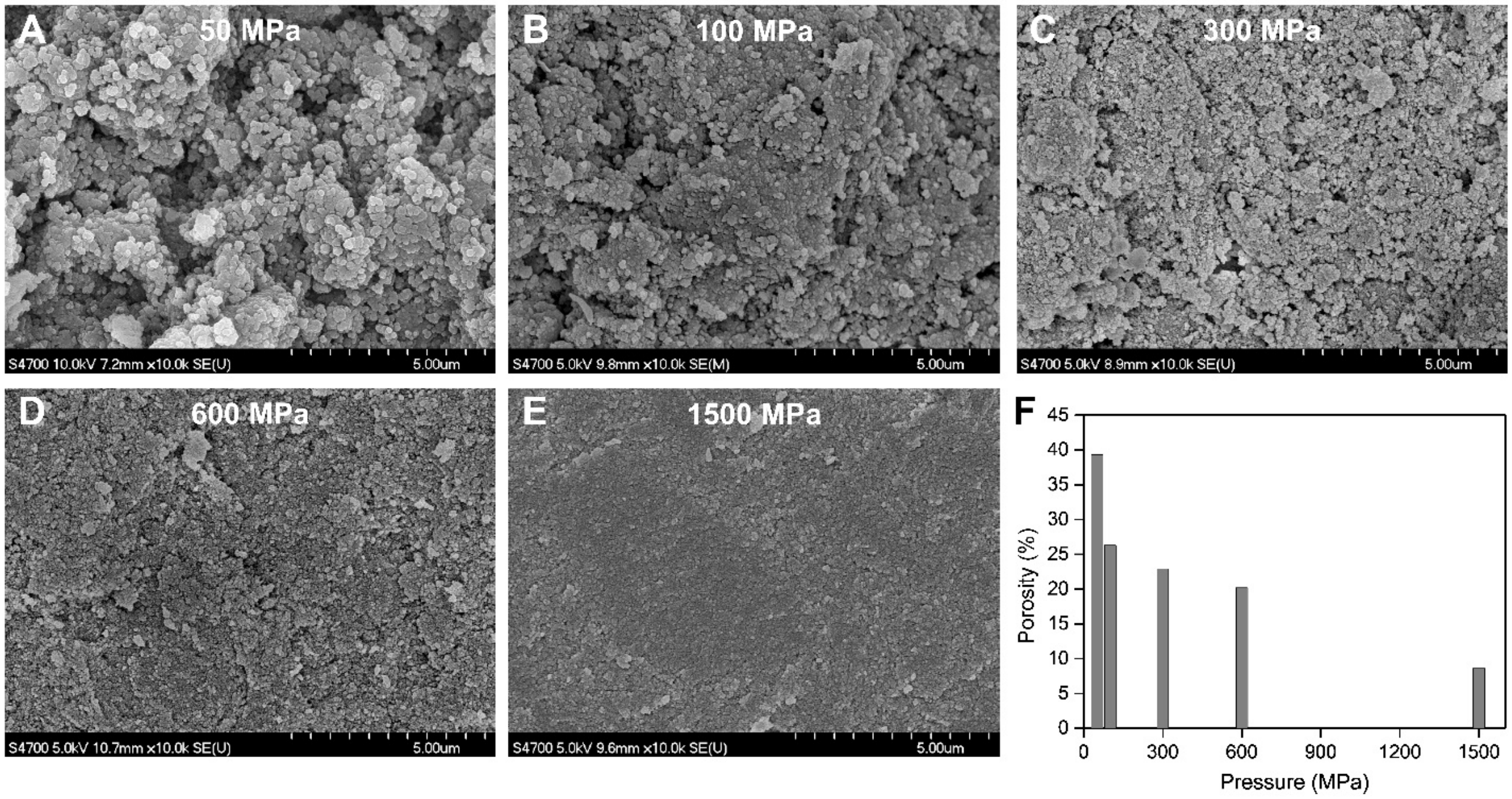

**Fig. S14.** SEM images of $Al_2O_3$ layer compacted with different pressures (**A**: 50 MPa, **B**:100 MPa, **C**: 300 MPa, **D**: 600 MPa, and **E**: 1500 MPa) and their corresponding porosities **(F)**.

**Fig. S15.**

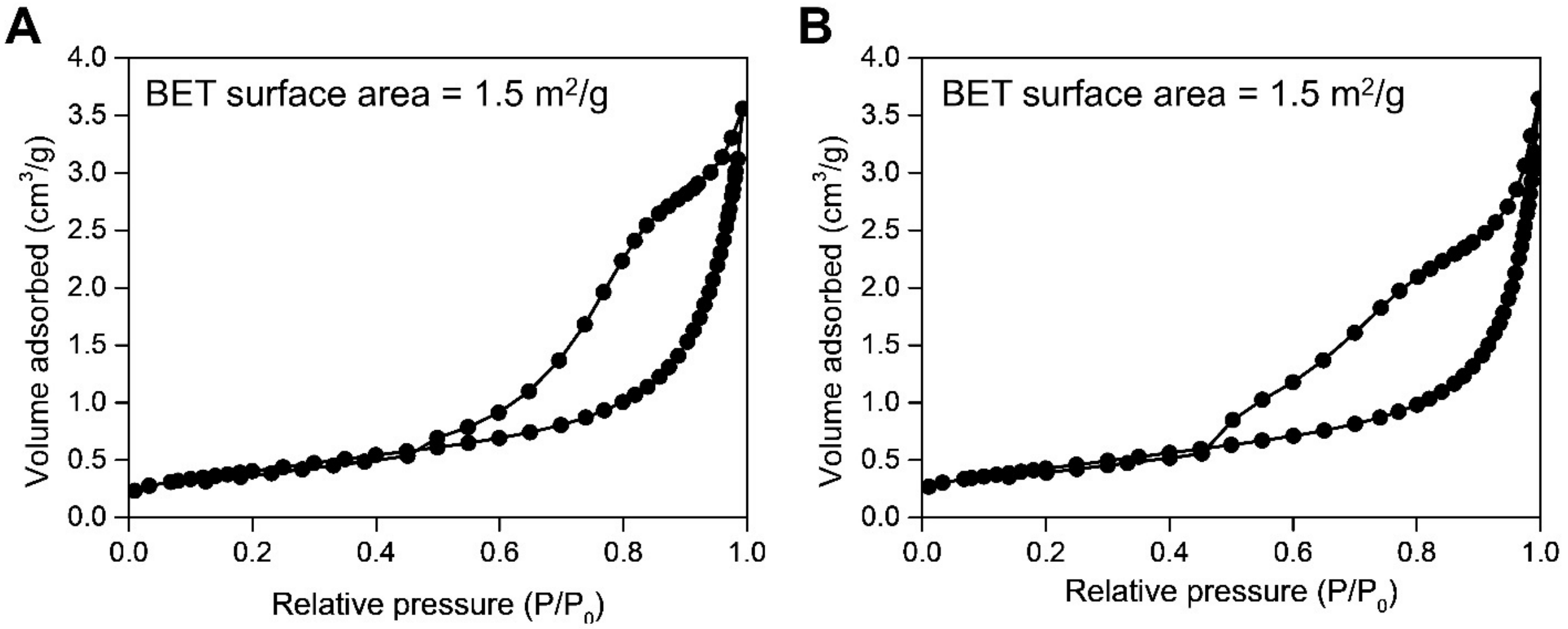

**Fig. S15.** $N_2$ adsorption-desorption isotherm curves and BET surface areas of **(A)** $LiNi_{0.8}Co_{0.15}Al_{0.05}O_2$ (LNCA) powder and **(B)** LNCA pellet (pressed at 100MPa).